\documentclass[preprint, a4paper]{revtex4}
\usepackage[T1]{fontenc}
\usepackage{graphicx}
\usepackage{psfrag}
\usepackage{rotating}
\usepackage{bm}        
\usepackage{amssymb}   
\usepackage{mathrsfs}
\usepackage{bm}
\def\ddpar{\partial}

\def\k1{k_1}
\def\k2{k_2}
\def\q1{q_1}
\def\q2{q_2}

\def\({\left (}
\def\){\right )}
\def\[{\left [}
\def\]{\right ]}

\newcommand{\beq}{\begin{equation}}
\newcommand{\eeq}{\end{equation}}

\newcommand{\dB}{\Delta B}
\newcommand{\aB}{a_{\mathrm{B}}}

\newcommand{\mILi}{M_{I_{\rm {Li}}}}
\newcommand{\mIRb}{M_{I_{\rm {Rb}}}}

\newcommand{\mSLi}{M_{S_{\rm {Li}}}}
\newcommand{\mSRb}{M_{S_{\rm {Rb}}}}

\begin{document}
\date{\today}
\title{
Effects of Electric Fields on Heteronuclear Feshbach Resonances in Ultracold $^6\rm{Li}-^{87}\rm{Rb}$ Mixtures}
\author{Z. Li$^1$, and K. W. Madison$^2$}
\affiliation{
$^{1}$Department of Chemistry, University of British Columbia, Vancouver, Canada, V6T 1Z1\\
$^{2}$Department of Physics and Astronomy, University of British Columbia, Vancouver, Canada V6T 1Z1}

\begin{abstract}
The effects of combined external electric and magnetic fields on elastic collisions in ultracold Li--Rb mixtures is studied using recently obtained, experimentally verified potentials. Our analysis provides  both quantitative predictions for and a detailed physical interpretation of the phenomena arising from electric-field-induced interactions. It is shown that the electric field shifts the positions of intrinsic magnetic Feshbach resonances, generates copies of resonances previously restricted to a particular partial-wave collision to other partial wave channels, and splits Feshbach resonances into multiple resonances for states of non-zero angular momenta.  It was recently observed that the magnetic dipole-dipole interaction can also lift the degeneracy of a $p$-wave state splitting the associated $p$-wave Feshbach resonance into two distinct resonances at different magnetic fields.  Our work shows that the splitting of the resonances produced by an applied electric field is more than an order of magnitude larger.  This new phenomenon offers a complementary way to produce and tune an anisotropic interaction and to study its effect on the many-body physics of heteronuclear atomic gases.
\end{abstract}

\maketitle

\section{Introduction}

The discovery of magnetic field tunable Feshbach resonances (FRs) has led to many groundbreaking experiments in the field of ultracold atomic and molecular physics \cite{Tiesinga,Weiner, Kohler}. Magnetic FRs provide a powerful tool to control microscopic interactions in ultracold quantum gases  \cite{Courteille,Inouye}, offer an extremely sensitive probe of interatomic interaction potentials for collisions at ultracold temperatures \cite{Leo,Li,Marzok}, and can be used to create ultracold molecules by coherently linking ultracold atoms \cite{Danzl, Ni}. FRs arise due to coupling between a quasi-bound molecular state in a closed collision channel and the scattering wave function of the colliding atoms in an open channel.  Because the quasi-bound states and the free atomic pair have, in general, different magnetic moments, both their absolute energy and relative energy difference can be tuned using an external magnetic field.  When the energy of the quasi-bound state is degenerate with the energy of the free atomic pair, a resonant scattering process occurs, the $s$-wave scattering length diverges and both elastic and inelastic collisions are dramatically enhanced. Recent theoretical work has also demonstrated the possibility of inducing FRs in heteronuclear mixtures of atomic gases by applying a static electric field \cite{Krems,Li2}.  The mechanism is based on the interaction of the instantaneous dipole moment of the heteronuclear collision complex with the external electric field.  This interaction is distinct from and, for bi-alkali mixtures with a large electric dipole moment, much larger than that responsible for electric field control of ultra-cold collisions based on the electric polarization of colliding atoms and the resulting dipole-dipole interaction \cite{Marinescu, Melezhik, Marcelis}.  The interaction considered here, for heteronuclear collisions, couples collision states of different orbital angular momenta, and the coupling becomes very significant near a FR. This coupling gives rise to new $s$-wave scattering resonances induced by the presence of FRs in higher partial wave states and can shift the positions of the quasi-bound states resulting in a shift in the positions of the intrinsic magnetic FRs (i.e.~those present in the absence of an external electric field).  Moreover, the electric field can induce a strong anisotropy of ultracold scattering by exerting a torque on the collision complex of ultracold atoms.

The use of combined electric and magnetic fields to control interatomic and intermolecular interactions has several distinct advantages over using magnetic fields alone.  It has been demonstrated that the combination of electric and magnetic fields may be used to control both the position and width of FRs independently even for homonuclear collisions, leading to complete control over the character of ultracold collisions \cite{Marcelis}.  In addition, for the relevant field strengths considered here ($\le 100$~kV/cm and $\le1$~kG), the electric fields can be varied much faster than the magnetic fields, in large part, because the corresponding field energy density is 10 times smaller.  Electric fields induce anisotropic interactions and angle-dependent scattering at ultra-cold temperatures, which may affect the dynamics of quantum degenerate gases in unanticipated ways.  It is also interesting to note that the electric-field control of heteronuclear collisions can be achieved at fields low enough that
they do not perturb the separated atoms or non-polar molecules since they interact significantly with electric field only when in a collision complex.  For these reasons, electric field control of interatomic interactions may be preferable to magnetically or optically tunable scattering resonances in certain applications.

Silber \emph{et al.} have recently created a quantum degenerate Bose-Fermi mixture of $^6\rm Li$ and $^{87}\rm Rb$ atoms in a magnetic trap \cite{Silber} and detected inter-species FRs \cite{Deh}.  These resonances may eventually be useful to improve the efficiency of sympathetic cooling in this mixture and for the study of boson-mediated BCS pairing \cite{Bijlsma}. These FRs also provide a method to create loosely bound LiRb dimers. LiRb molecules have a relatively large electric dipole moment (up to 4.2 Debye) \cite{Aymar}, which makes the Li--Rb system a good candidate for research on ultracold dipolar gases and the experimental study of electric-field-induced FRs \cite{Krems,Li2}. Recently, we have generated accurate singlet and triplet interaction potentials for ultracold collisions in $^6$Li--$^{87}$Rb gaseous mixtures by fitting the experimentally measured FRs \cite{Li}. In this work, we use these potentials to investigate in detail the effects of combined external electric and magnetic fields on elastic collisions in ultracold Li--Rb mixtures. To guide future experimental studies, we predict the positions and widths of electric-field-induced FRs for several spin states and explore the effect of the orientation of the electric field with respect to the magnetic field on ultracold elastic collisions.  The work presented here represents the first quantitative analysis of electric-field-induced resonances based on precise inter-atomic potentials.  In addition, our analysis provides insights into the detailed physical mechanism of electric-field-induced interactions in ultracold binary mixtures of alkali metal atoms.  We report for the first time the observation that the coupling induced by electric fields splits FRs into multiple resonances for states of non-zero angular momenta. Recently it was observed that the magnetic dipole-dipole interaction can lift the degeneracy of a $p$-wave state \cite{Regal}.  This splits a $p$-wave FR into two distinct resonances at different magnetic fields \cite{Regal,Ticknor}.  The splitting of the resonances studied here is produced only in heteronuclear collisions by a coupling between different quasi-bound states, is continuously tunable using an applied electric field, and is more than an order of magnitude larger than the splitting induced by magnetic dipole-dipole interactions.  This new phenomenon offers a complementary way to produce and tune an anisotropic interaction and to study its effect on the many-body physics of heteronuclear atomic gases.  

\section{Theory}

The Hamiltonian for the $^6$Li--$^{87}$Rb collision system (or any bi-alkali) in the presence of superimposed electric and magnetic fields is 
\begin{equation}
\hat H = \hat H_{\rm rel} +\hat{V}_E(R)+ \hat{V}_B + \hat {V}_{\rm hf}
\end{equation}
where $\hat H_{\rm rel}$ accounts for the relative motion of the atoms, $\hat{V}_E(R)$ describes the interaction between LiRb and an external electric field, $\hat{V}_B$ models the interaction of the collision complex with an external magnetic field, and $\hat {V}_{\rm hf}$ represents the hyperfine interactions. $\hat H_{\rm rel}$ can be written as
\begin{eqnarray}
\hat H_{\rm rel} = - \frac{1}{2\mu R}
\frac{\ddpar^2}{\ddpar R^2}{R} 
+\frac{\hat{l}^2(\theta,\phi)}{2\mu R^2}+\hat{V}(R)
\end{eqnarray}
where \begin{math}\mu \end{math} is the reduced mass of the colliding atoms, \begin{math}R\end{math} is the interatomic distance, \begin{math}\hat l\end{math} is the operator describing the rotation of the collision complex and the angles \begin{math}\theta\end{math} and \begin{math}\phi\end{math} specify the orientation of the interatomic axis in the space-fixed coordinate frame. Here, we neglect the magnetic dipole-dipole interaction since it has a negligible effect on the observables described in this paper. The atomic and molecular quantum numbers used in this article are defined in Table~\ref{tab:quantumnumbers}.

We expand the total wave function of the diatomic system in a fully uncoupled, space-fixed basis set:
\begin{equation}
\psi =\frac{1}{R} \sum_{\alpha}\sum_l \sum_{m_l} F_{\alpha l m_{l}}(R) \,
| l m_{l}\rangle  \; | \alpha \rangle
\end{equation}
where $F_{\alpha l m_{l}}(R) \, |l m_{l}\rangle$ and $|\alpha\rangle=| I_{\rm Li} M_{ I_{\rm Li}}\rangle| S_{\rm Li} M_{S_{\rm Li}}\rangle | I_{\rm Rb} M_{ I_{\rm Rb}}\rangle  | S_{\rm Rb} M_{S_{\rm Rb}}\rangle$ are the radial basis states and the atomic spin states, respectively.
The substitution of this expansion in the Schr\"odinger equation with the Hamiltonian (1) results in a system of coupled differential equations
\begin{eqnarray}
\left[ \frac{d ^2}{d R^2} - \frac{l(l+1)}{R^2} + 2\mu \epsilon \right] F_{\alpha lm_{l}}(R)=
\nonumber\\
2\mu  \sum_{\alpha'}\sum_{l'}\sum_{m'_l} \left \langle \alpha l m_l  \left| \hat V(R)+\hat{V}_E(R)+ \hat{V}_B + \hat{V}_{\rm hf} \right| \alpha' l' m'_l\right\rangle F_{\alpha' l'm'_{l}}(R)
\label{coupled-equations}
\end{eqnarray}
which we solve at fixed values of total energy \begin{math}\epsilon\end{math}.

The electronic interaction potential \begin{math}\hat{V}(R)\end{math} can be represented as
\begin{eqnarray}
\hat {V}(R)=\sum_{S}\sum_{M_S}| SM_S\rangle{V}_S(R)\langle SM_S|
\end{eqnarray}
where \begin{math}V_S(R)\end{math} denotes the adiabatic interaction potential of the molecule in the spin state \begin{math}S\end{math}. 
To evaluate the matrix elements of the interaction potential \begin{math}\hat V(R)\end{math}, we write the atomic spin states \begin{math}| I_{\rm Li} M_{ I_{\rm Li}}\rangle| S_{\rm Li} M_{S_{\rm Li}}\rangle | I_{\rm Rb} M_{ I_{\rm Rb}}\rangle  | S_{\rm Rb} M_{S_{\rm Rb}}\rangle\end{math} in terms of the total electronic spin
\begin{eqnarray}
| I_{\rm Li} M_{ I_{\rm Li}}\rangle| S_{\rm Li} M_{S_{\rm Li}}\rangle | I_{\rm Rb} M_{ I_{\rm Rb}}\rangle  | S_{\rm Rb} M_{S_{\rm Rb}}\rangle
=
\nonumber\\
\sum_S \sum_{M_S} 
(-1)^{M_S} (2S + 1)^{1/2}
\left ( \begin{array}{ccc}
S_{\rm Li} & S_{\rm Rb} & S \\
M_{S_{\rm Li}}&M_{S_{\rm Rb}}& -M_S
\end{array}
\right ) 
| I_{\rm Li} M_{ I_{\rm Li}}\rangle | I_{\rm Rb} M_{ I_{\rm Rb}}\rangle  | S M_S\rangle
\label{recoupling}
\end{eqnarray}
and note that
\begin{eqnarray}
\langle S M_S|\hat V(R)| S' M'_S\rangle=V_S(R)\ \delta_{SS'}\delta_{M_S M'_S}.
\end{eqnarray}
The term enclosed in parentheses in Eq.~\ref{recoupling} denotes a 3$j$-symbol.
The operator \begin{math}\hat V(R)\end{math} is diagonal in the nuclear spin states and \begin{math}l\end{math} and \begin{math}m_l\end{math} quantum numbers.

The operator \begin{math}\hat{V}_E(R)\end{math} can be written in the form
\begin{eqnarray}
\hat{V}_E(R) = - \vec{E} \cdot \vec{d} = -E (\hat{e}_E \cdot \hat{e}_d) \sum_{S}\sum_{M_S}| SM_S\rangle d_S(R)\langle SM_S|
\label{E-field}
\end{eqnarray}
where $\hat{e}_E$ and $\hat{e}_d$ are the unit vectors pointing along the electric field and dipole moment of the LiRb dimer respectively, $d_S$ denotes the spin dependent dipole moment functions of LiRb, and $E$ is the electric field magnitude.  Clearly, the electric field coupling depends on the orientation of the field with respect to the dipole moment.  Specifically, $\hat{e}_E \cdot \hat{e}_d = \cos (\chi)$ where $\chi$ is the angle between $\vec{E}$ and $\vec{d}$.
If the electric field and dipole moment vectors are oriented at angles $\gamma$ and $\theta$ with respect to the quantization axis (taken to be along the $\hat{z}$-axis), then this term can be written in terms of the first-degree Legendre polynomial as 
\begin{eqnarray}
\hat{e}_E \cdot \hat{e}_d = \cos(\chi) = P_1(\cos(\chi)) \\\nonumber
= \frac{4\pi}{3}\[Y_{1 -1}^*(\gamma,\phi_\gamma)Y_{1 -1}(\theta,\phi_\theta) + Y_{1 0}^*(\gamma,\phi_\gamma)Y_{1 0}(\theta,\phi_\theta) + Y_{1 1}^*(\gamma,\phi_\gamma)Y_{1 1}(\theta,\phi_\theta)\]
\end{eqnarray}
where the $Y_{\rm x x}$ are spherical harmonics and
we have used the spherical harmonic addition theorem.  The azimuthal angles, $\phi_\gamma$ and $\phi_\theta$, are measured from the positive $x$-axis to the orthogonal projection of the $\vec{E}$ and $\vec{d}$ vectors in the $x$-$y$ plane. Figure \ref{coordinate} illustrates this coordinate system.

The dipole moment functions are modeled by
\begin{equation}
d_S(R) = \rm{D} \exp{\left [ - {\alpha (\rm{R}-\rm{R_e})^2}\right]}
\label{dipolefunction}
\end{equation}

\noindent
with the parameters $R_e = 7.2 \; \aB$, $\alpha = 0.06 \; \aB^{-2}$ and $D = 4.57$~Debye for the singlet state, and $R_e = 5.0 \; \aB$, $\alpha = 0.045 \; \aB^{-2}$ and $D = 1.02$~Debye for the triplet state, where the Bohr radius is $\aB = 0.0529177$~nm.  These analytical expressions approximate the true functions are were fit to the numerical data for the dipole moment functions computed by  Aymar and Dulieu \cite{Aymar}.

The collision dynamics in Li--Rb system depends on the relative angle ($\gamma$) between the electric and magnetic fields. Without loss of generality, we can assume that $\vec E$ lies in the $x$-$z$ plane, i.e., $\phi_\gamma = 0$. The matrix elements of $\hat{V}_E(R)$ are therefore evaluated using the expressions
\begin{eqnarray}
\label{electric}
\langle l m_l| 
\hat{e}_E \cdot \hat{e}_d | l' m'_l\rangle=
\frac{1}{\sqrt{2}} \sin{\gamma} (-1)^{m_l'}\sqrt{(2l+1)(2l'+1)}
\left(\begin{array}{ccc}
l&1&l'\\
0&0&0\\
\end{array}\right)
\left(\begin{array}{ccc}
l&1&l'\\
m_l&-1&-m_l'\\
\end{array}\right)\\\nonumber
+ \cos{\gamma} (-1)^{m_l'}\sqrt{(2l+1)(2l'+1)}
\left(\begin{array}{ccc}
l&1&l'\\
0&0&0\\
\end{array}\right)
\left(\begin{array}{ccc}
l&1&l'\\
m_l&0&-m_l'\\
\end{array}\right)\\\nonumber
- \frac{1}{\sqrt{2}} \sin{\gamma} (-1)^{m_l'}\sqrt{(2l+1)(2l'+1)}
\left(\begin{array}{ccc}
l&1&l'\\
0&0&0\\
\end{array}\right)
\left(\begin{array}{ccc}
l&1&l'\\
m_l&1&-m_l'\\
\end{array}\right)
\end{eqnarray}
and
\begin{eqnarray}
\langle SM_S| \left( \sum_{S''} \sum_{M''_S}| S'' M''_S\rangle d_{S''}\langle S'' M''_S| \right) | S' M'_S\rangle
=d_S\ \delta_{SS'}\delta_{M_S M'_S}.
\end{eqnarray}
It is important to note that
the electric field coupling operator, $\hat{V}_E(R)$,
has a spin structure identical to the electronic interaction potential, $\hat{V}(R)$.
Namely, they are both diagonal in the total electronic spin and its $z$ projection.  However,
as is clear from Eq.~\ref{electric},
the geometric factor, $\hat{e}_E \cdot \hat{e}_d$, 
introduces this additional electric-field-induced coupling only between states of \emph{different} orbital angular momenta.
The first 3$j$-symbol in Eq.~\ref{electric} is non-zero for $l+l' = \pm1$.  The second 3$j$-symbol in Eq.~\ref{electric} selects states with orbital angular momenta projections
differing by $\Delta m_l = m_l - m_l' = \pm1,\; 0$.
If the electric field is directed along the $z$-axis (the quantization axis) then $\gamma = 0$ and Eq.~{\ref{electric} reduces to
\begin{eqnarray}
\langle l m_l| \cos{\theta} | l' m'_l\rangle=
\delta_{m_l m_l'}(-1)^{m_l} \sqrt{(2l+1)(2l'+1)}
\left(\begin{array}{ccc}
l&1&l'\\
-m_l&0&m_l\\
\end{array}\right)
\left(\begin{array}{ccc}
l&1&l'\\
0&0&0\\
\end{array}\right).
\end{eqnarray}
In this case only those states with the same orbital angular momenta projections are coupled.

In the absence of an electric field, the Hamiltonian (neglecting magnetic dipole-dipole interactions) preserves the projection of the total angular momentum along the quantization (magnetic field) axis, i.e.~the sum $\mSLi + \mSRb + \mILi + \mIRb + m_l$ is conserved.  In addition, the orbital angular momentum is also conserved. Therefore, the intrinsic ($E=0$) FRs which are present for a particular atomic spin state arise from its coupling to the set of bound states with the same total angular momentum projection and orbital angular momentum.  The energy of the bound states relative to the threshold state and their relative magnetic moments determine the magnetic fields at which resonances will occur and the strength of the coupling to a particular bound state determines the width of the resulting resonance.  The electric field induces additional couplings between the threshold state and bound states of different angular momenta but still within the same set of states sharing the same total spin angular momentum projection $m_F = \mSLi + \mSRb + \mILi + \mIRb$.  The result is that FRs previously restricted to a particular partial-wave collision will appear on adjacent partial wave scattering states.  The widths of these electric-field-induced FRs depend on the strengths of the couplings and therefore on the magnitude of the electric field, $E$.  In addition, the electric field coupling among the bound states (with the same total $m_F$ value) will shift their absolute energies and therefore the positions of the intrinsic FRs.  Moreover, this coupling and the resulting shifts are, in general, dependent on $m_l$ and, as a consequence, resonances associated with bound states with $l>0$ will split into $l+1$ distinct resonances.

The interaction of the atoms with an external magnetic field \begin{math} B\end{math} is described by
\begin{eqnarray}
\hat {V}_B = 2\mu_0 B \left( \hat{S}_{Z_{\rm Li}} + \hat{S}_{Z_{\rm Rb}} \right)
- B \left( \frac{\mu_{\rm Li}}{I_{\rm Li}}\hat{I}_{Z_{\rm Li}}
+\frac{\mu_{\rm Rb}}{I_{\rm Rb}}\hat{I}_{ Z_{\rm Rb}} \right)
\end{eqnarray}
where $B$ is the magnetic field strength (directed along the $z$-axis),   \begin{math}\mu_0\end{math} is the Bohr magneton and \begin{math}\mu_{\rm Li (Rb)}\end{math} denote the nuclear magnetic moments of Li (Rb). $\hat{S}_{Z_{\rm Li (Rb)}}$ and $\hat{I}_{Z_{\rm Li (Rb)}}$ give the $z$ component of the operators describing the electronic and nuclear spins of Li (Rb), $\hat{S}_{\rm Li (Rb)}$ and $\hat{I}_{\rm Li (Rb)}$, respectively. The hyperfine interaction \begin{math}\hat {V}_{\rm hf}\end{math} can be represented as
\begin{eqnarray}
\hat {V}_{\rm hf} = \gamma_{\rm Li} \hat{I}_{\rm Li} \cdot \hat{S}_{\rm Li}
+\gamma_{\rm Rb}\hat{I}_{\rm Rb}\cdot \hat{S}_{\rm Rb}
\end{eqnarray}
where \begin{math}\gamma_{\rm Li}\end{math} and \begin{math}\gamma_{\rm Rb}\end{math} are the atomic hyperfine interaction constants: $\gamma_{\rm Li}= 152.14$ MHz and $\gamma_{\rm Rb}= 3417.34$~MHz (where we work in units with $\hbar = 1$).

The operator representing the magnetic field interaction is diagonal in the representation \begin{math}| I_{\rm Li} M_{ I_{\rm Li}}\rangle| S_{\rm Li} M_{S_{\rm Li}}\rangle | I_{\rm Rb} M_{ I_{\rm Rb}}\rangle  | S_{\rm Rb} M_{S_{\rm Rb}}\rangle\end{math}, while the matrix elements of the hyperfine interaction operators are not but can be readily evaluated using the relations:
\begin{equation}
\hat I_{\rm Li} \cdot \hat S_{\rm Li}=\hat{I}_{Z_{\rm Li}} \hat{S}_{Z_{\rm Li}} + \frac{1}{2}(\hat{I}_{{\rm Li}+}\hat{S}_{{\rm Li}-}+\hat{I}_{{\rm Li}-}\hat{S}_{{\rm Li}+})
\end{equation}
and
\begin{equation}
\hat I_{\rm Rb} \cdot \hat S_{\rm Rb}=\hat{I}_{Z_{\rm Rb}} \hat{S}_{Z_{\rm Rb}} + \frac{1}{2}(\hat{I}_{{\rm Rb}+}\hat{S}_{{\rm Rb}-}+\hat{I}_{{\rm Rb}-}\hat{S}_{{\rm Rb}+}).
\end{equation}
$\hat{I}_{\pm}$ and $\hat{S}_{\pm}$ are the raising and lowering operators. Because of these terms, the matrix of the Hamiltonian in the basis \begin{math}| \alpha l m_{l}\rangle\end{math} does not become diagonal as \begin{math}R \to \infty \end{math}. Therefore, we find the basis which diagonalizes the matrix of $\hat{V}_E+\hat{V}_B+\hat V_{\rm hf}$, and transform the solutions of Eq.~\ref{coupled-equations} into this basis.  At this point, the boundary conditions are applied and we construct the scattering $S$-matrix.  This procedure has been described in Ref. \cite{Krems3}. The scattering matrix thus obtained yields the probabilities of elastic and inelastic scattering of Li and Rb in the presence of electric and magnetic fields.

\section{results}

The Li--Rb mixture is an important system for the study of both ultracold atomic and molecular gases. A quantum degenerate Bose-Fermi mixture of $^6$Li and $^{87}$Rb atoms has been recently created \cite{Silber} and may be important for research of ultracold fermionic and bosonic mixtures. By tuning an external magnetic field through a FR, researchers can create an ensemble of LiRb molecules, a polar dimer, from this gaseous mixture.  Since LiRb molecules have a large electric dipole moment, the Li--Rb system is also a good candidate for the research of ultracold dipolar gases and the experimental study of electric-field-induced FRs \cite{Li2}.  Motivated by these features and the availability of accurate potentials \cite{Li}, we analyze the collision properties of Li--Rb mixtures in the presence of both magnetic and electric fields.

As discussed above, in the absence of electric fields, different partial wave states $|l m_l\rangle$ of the Li--Rb collision complex are uncoupled and $s$-wave scattering entirely determines the collision dynamics in ultracold Li--Rb gases. The presence of an external electric field, however, induces couplings between states of different orbital angular momenta with $\Delta l = \pm1$.  As a result, a resonant enhancement of the $s$-wave cross section appears at magnetic fields near intrinsic $p$-wave resonances.  
Figure~\ref{fig:ElectricFieldInducedWholePicture} shows the magnetic field dependence of $s$- and $p$-wave elastic cross sections for Li and Rb atoms in the spin state $|\frac{1}{2}, \frac{1}{2} \rangle_{^6\rm Li} \otimes |1, 1 \rangle_{^{87}\rm Rb}$ (where $|f, m_f\rangle$ is the usual notation for the atomic hyperfine states) computed at zero electric field and at $E = 100$~kV/cm.  Here the electric field is directed along the quantization axis ($\gamma=0$).  In the presence of the 100~kV/cm electric field, an $s$-wave resonant peak appears (indicated at $A$) at the magnetic field of 877.50~G arising from an intrinsic $p$-wave resonance. We refer to this resonance as an \emph{electric-field-induced Feshbach resonance}.  Figure \ref{fig:s-wave-induced-resonance} shows this feature in more detail.  We present in Table 
\ref{tab:resonances} the positions and widths of electric-field-induced FRs for several atomic spin states of the $^6$Li--$^{87}$Rb system at magnetic fields below 2~kG.  For each resonance, we extract the position ($B_0$) and width ($\dB$) from the magnetic field dependence of the scattering length.
\begin{equation}
a(B) = a_{\rm bg} \(1 - \frac{\dB}{B_0 - B}\)
\end{equation}
where $a_{\rm bg}$ represents the background scattering length. In this calculation, we also observe new $p$-wave resonances induced by the coupling to a $d$-wave state, and we find that these $p$-wave resonances give rise to new $s$-wave electric-field-induced FRs, denoted by ($d$) in Table~\ref{tab:resonances}.

The width of the electric-field-induced FRs is determined by the strength of the coupling, which is, in turn determined by the magnitude of the electric field.  In Fig.~\ref{fig:width-vs-E}, we plot the width of the $s$-wave resonance (shown in Fig.~\ref{fig:s-wave-induced-resonance}) induced by the intrinsic $p$-wave resonance near 882~G as a function of the electric field magnitude.  We find that the width can be well represented by a quadratic function of $E$, at least for the electric fields below 200~kV/cm, which suggests that this induced resonance arises from an indirect coupling \cite{hemming:022705}.

The electric field not only induces new resonances but also shifts the position of intrinsic magnetic FRs. Figure \ref{fig:ElectricFieldInducedWholePicture} shows that the interaction of Li--Rb dipole moment with the electric field shifts the position of both the $s$- and $p$-wave resonances. 
At $B$ and at $C$ an intrinsic $s$-wave resonance is shifted to higher magnetic fields (corresponding to a shift of the associated bound state to lower energy) due to the electric field.  At $D$ an intrinsic $p$-wave resonance is shifted to lower magnetic fields (corresponding to a shift of the associated bound state to higher energy).  For the most part, the shift of the FR positions arises from the coupling between bound states whereas the coupling of a given bound state to the scattering state results in a broadening of the associated electric-field-induced resonance.  We note that the shift of magnetic FRs at higher magnetic fields (e.g.~$C$) is more significant than the shift at lower magnetic fields  (e.g.~$B$). This generic behavior results from the fact that resonances associated with higher magnetic fields are typically more deeply bound than those associated with lower magnetic fields.  As a result, the wave function of the bound state giving rise to FRs at higher fields samples smaller interatomic distances where the dipole moment function is much larger.

Another example of the shift induced by the electric field couplings is shown in Fig.~\ref{fig:gradual-shift} for atoms in the atomic spin state $|\frac{1}{2}, -\frac{1}{2} \rangle_{^6\rm Li} \otimes |1, -1 \rangle_{^{87}\rm Rb}$.  An electric field of 30~kV/cm is large enough to shift the position of this $s$-wave resonance by almost 2~G - much larger than its width - while a field of 100~kV/cm produces a shift of almost 9~G.  It is important to note that this $s$-wave resonance shifts to lower magnetic fields as the electric field increases, and this is opposite to the shift of the $s$-wave resonances shown in Fig.~\ref{fig:ElectricFieldInducedWholePicture}.  The shift of a resonance results from level repulsion between the closed channel bound states and therefore depends on the proximity, position, and coupling strengths of the nearby bound states.  Therefore, the direction of the resonance shift and its dependence on the electric field magnitude do not exhibit a generic behavior but depend on the particular environment of a given resonance.

These shifts provide a way to dramatically and rapidly modify the $s$-wave scattering length by tuning into and out of an intrinsic magnetic field resonance. Fig.~\ref{fig:CrossSectionVsElectricField} presents two such tuned resonances arising from the variation of the electric field for atoms in the atomic spin state $|\frac{1}{2}, \frac{1}{2} \rangle_{^6\rm Li} \otimes |1, 1 \rangle_{^{87}\rm Rb}$. This figure shows the cross section for $s$-wave collisions as a function of the electric field strength with the magnetic field fixed at 1066~G (solid line) and 878~G (dotted line).  The solid curve shows a large resonance feature due the intrinsic magnetic FR at 1067~G which shifts to higher magnetic field (lower absolute energy) as the electric field increases (see Fig.~\ref{fig:ElectricFieldInducedWholePicture}).  The small resonance feature which appears at the electric field strength of approximately 16~kV/cm in the solid curve arises from an electric-field-induced resonance arising from the intrinsic $p$-wave resonance just above 1066~G which shifts to lower magnetic field (higher energy) as the electric field increases.  In the same plot, the dotted curve shows a resonance feature due to the shift of an electric-field-induced resonance arising from the intrinsic $p$-wave resonance at 882~G.  Fig.~\ref{fig:s-wave-induced-resonance} shows the $p$-wave state responsible for this resonance shifts to lower magnetic fields (higher energy) as the electric field increases.

In the presence of an electric field, the couplings between different partial wave states can push the bound states in, for example, the $s$-wave and $p$-wave interaction potentials apart.  This level repulsion gives rise to the electric-field-induced shift of the intrinsic $s$- and $p$-wave magnetic FRs.  Since the couplings depend on the orbital angular momentum projection, $m_l$, we also expect the electric field induced coupling to also \emph{split} FRs for states of non-zero angular momenta. This mechanism is illustrated in Fig.~\ref{fig:ResonanceRepulsionFig} where three adjacent bound state levels are shown as well as the coupling induced by an applied electric field with $\gamma=0$. Without external electric fields, the bound states in the $p$-wave interaction potential are degenerate, whereas the electric field lifts this degeneracy.  In the case where the electric field points along the quantization axis ($\gamma =0$), the $m_l=0$ bound state in the $p$-wave potential is coupled to bound states in both the $s$- and $d$-wave potentials whereas the $|m_l|=1$ bound states are only coupled to bound states in the $d$-wave potential.  This occurs because the system is cylindrically symmetric and the couplings between internal spin states and partial wave states are negligible.  As a result, the coupled states repel and the $m_l=0$ state is shifted differently than the $|m_l|=1$ states splitting the $p$-wave resonance into a doublet. For the purposes of simplifying the discussion, we have neglected the coupling to yet higher order partial wave states and we have neglected the possible presence of other closed channel states in the near vicinity. This mechanism generally applies to all nonzero partial waves. For a state with an orbital angular momentum $l$, the number of peaks is $l +1$ corresponding to the number of distinct values for $|m_l|$.  Fig.~\ref{fig:p-splitting} and Fig.~\ref{fig:d-splitting} show the splitting of a $p$-wave and a $d$-wave FR, respectively.  In the presence of a 100~kV/cm electric field, the $p$-wave resonance splits into two peaks (corresponding to the $|m_l| = 1$ and $m_l = 0$ components) with a separation of 4~G (dash-dot line in Fig.~\ref{fig:p-splitting}). The shift of the $|m_l| = 1$ peak in Fig.~\ref{fig:p-splitting} to higher magnetic fields (lower energy) is consistent with coupling between the $p$-wave bound states and a $d$-wave state which resides at a higher energy (illustrated in Fig.~\ref{fig:d-splitting}).  The splitting of a $d$-wave bound state gives rise to three separated resonances and is shown in Fig.~\ref{fig:d-splitting}.  An interval of 1~G opens up between $m_l = 0$ and $|m_l| = 1$ and the an interval of 2~G appears between $|m_l| = 1$ and $|m_l| = 2$. Since it is only very weakly coupled to higher partial-wave states, the $|m_l| = 2$ component remains in essentially the same location as the resonance at zero electric field.

The splitting of FRs in states with non-zero orbital angular momenta has been previously discovered in experiments by Jin and coworkers \cite{Regal} and studied theoretically by Ticknor \emph{et al.} \cite{Ticknor}. They found that $p$-wave FRs for collisions of homonuclear gases of $^{40}$K split into a doublet due to the magnetic dipole-dipole interaction.  In the work presented here, we neglect the magnetic dipole-dipole interaction since it produces a negligible effect compared to the electric field coupling and the splitting we predict for FRs is entirely due to the effect of the electric field.  As discussed in \cite{Regal}, the ability to introduce and tune an anisotropic interaction using high-partial-wave resonances may have far reaching consequences for the study of novel forms of superfluidity using cold atomic gases \cite{Tsuei}.

The splitting of the nonzero-partial-wave resonances arising from magnetic dipole-dipole interactions is very small and will disappear as the resonance becomes broad with increasing temperature. In contrast, the splitting observed here, occurring for heteronuclear atomic mixtures, is more than an order of magnitude larger.  This new phenomenon offers a complementary way to produce and tune an anisotropic interaction and to study its effect on the many-body physics of heteronuclear atomic gases.

So far, we have discussed the modifications of FRs induced by the application of an electric field parallel to the magnetic field ($\gamma=0$).  In addition, we study the effect of non-parallel fields ($\gamma \ne 0$).  In Fig.~\ref{fig:Erotation}, we show the variation of the total elastic cross section for $p$-wave collisions given fixed electric (100~kV/cm) and magnetic fields as a function of the angle between them $\gamma$.  In the upper panel, the magnetic field was 877~G which is near the $p$-wave resonance for the $m_l=0$ component (see Fig.~\ref{fig:p-splitting}).  Whereas, in the lower panel, the magnetic field was 881.9~G and falls in between the $m_l=0$ and $|m_l|=1$ resonances in the $p$-wave doublet.  In the latter case, the variation of the cross sections as a function of $\gamma$ is only a factor of 10, while at a magnetic field near one of the resonances, the cross section varies by almost 4 orders of magnitude as $\gamma$ changes by less than $30^{\circ}$.

Fig.~\ref{fig:gamma45} presents the magnetic field dependence of the total elastic cross section for different components of $p$-wave scattering at $\gamma=45^{\circ}$ near the intrinsic $p$-wave resonance at 882~G.  In this case, because the electric field couples states of differing $m_l$ values, the doublet structure of the $p$-wave resonance appears on each of the three $m_l$ components of the open channel.  This is in contrast to the case with $\gamma=0$ shown in Fig.~\ref{fig:p-splitting} where the coupling is only between states with the same $m_l$ value and each component exhibits a single resonance.  It should be clarified here that the electric-field-induced $s$-wave resonance arising from this $p$-wave resonance exhibits only the single resonance corresponding to the $m_l=0$ component of the $p$-wave bound state.  This is because (neglecting the magnetic dipole-dipole interaction) the orbital angular momentum projection along the \emph{electric-field axis} is conserved by the Hamiltonian, and $m_l=0$ for $s$-wave collisions in all coordinate frames.  On the other hand, a state with orbital angular momentum $l$ and projection $m_l$ defined with respect to the magnetic field axis will be a linear combination of states with all possible values of $m_l$ when represented with respect to the electric field axis \cite{Zare}.

Fig.~\ref{fig:different-gamma} presents the magnetic field dependence of the average $p$-wave elastic scattering cross section (averaged over all three components) for atoms in the spin state $|\frac{1}{2}, \frac{1}{2} \rangle_{^6\rm Li} \otimes |1, 1 \rangle_{^{87}\rm Rb}$ at $E=100$~kV/cm and with three orientations of the electric field, $\gamma$ = $0^{\circ}$, $45^{\circ}$, and $90^{\circ}$. The main point of this plot is to illustrate that the position of the resonances remains unchanged for different values of $\gamma$.  This is particularly important for the experimental search for these effects since it means that any variation of the orientation of the electric and magnetic fields does not adversely affect the visibility of these multiplet features.  Consequently, any inhomogeneities in the direction of the electric field over the confinement size of the atomic ensemble would also not affect their visibility.  Of course, since the positions of the resonances do depend on the electric field strength, any inhomogeneities in the magnitude of the electric field would result in inhomogenous broadening of the observed resonances.

\section{Conclusions}

We have presented quantitative predictions of the effects of combined external electric and magnetic fields on elastic collisions in ultracold Li--Rb mixtures.  This work is the first analysis of electric-field-induced interactions based on precise, experimentally verified inter-atomic potentials.  In addition, we have provided important insights into the detailed physical mechanism of electric-field-induced interactions in ultracold binary mixtures of alkali metal atoms.  We have shown that the electric field both shifts the position of intrinsic FRs and generates copies of resonances previously restricted to a particular partial-wave collision to other partial wave channels.  To facilitate the experimental search for these phenomena, we have provided predictions for the positions and widths of electric-field-induced FRs for several spin states and we have analyzed the effects of a non-parallel orientation of the electric field with respect to the magnetic field on ultracold elastic collisions.  We also have reported for the first time the observation that the coupling induced by electric fields splits FRs into multiple resonances for states of non-zero angular momenta.  It was recently observed that the magnetic dipole-dipole interaction can also lift the degeneracy of a $p$-wave state splitting the associated $p$-wave FR into two distinct resonances at different magnetic fields \cite{Regal,Ticknor}.  The primary differences with that work are that the splitting studied here is produced only in heteronuclear collisions, is continuously tunable using an applied electric field, and is more than an order of magnitude larger than the splitting induced by magnetic dipole-dipole interactions.  We believe the additional degrees of control offered by electric-field interactions will play an important role in future experiments on the many-body physics of heteronuclear atomic gases.

\begin{acknowledgments}
We thank Roman Krems for simulating discussions and for his careful reading of this manuscript.  This work was supported by the Natural Sciences and Engineering Research Council of Canada (NSERC), the Canadian Foundation for Innovation (CFI), and the Canadian Institute for Advanced Research (CIfAR).
\end{acknowledgments}

\newpage

\begin{table}
\caption{Definition of quantum numbers used in this paper.}
\vspace{0.5cm}
\begin{ruledtabular}
\begin{tabular}{l l @{} l}
$l$&orbital angular momentum of the diatomic system\\
$m_l$&projection of $l$ on the space-fixed quantization axis\\
$S$&total electronic spin angular momentum of the diatomic system\\
$M_S$&projection of $S$ on the space-fixed quantization axis\\
$I_{\rm {Li}}$&nuclear spin angular momentum of Li\\
$M_{I_{\rm {Li}}}$&projection of $I_{\rm {Li}}$ on the space-fixed quantization axis\\
$I_{\rm {Rb}}$&nuclear spin angular momentum of Rb\\
$M_{I_{\rm {Rb}}}$&projection of $I_{\rm {Rb}}$ on the space-fixed quantization axis\\
$S_{\rm {Li}}$&electronic spin angular momentum of Li\\
$M_{S_{\rm {Li}}}$&projection of $S_{\rm {Li}}$ on the space-fixed quantization axis\\
$S_{\rm {Rb}}$&electronic spin angular momentum of Rb\\
$M_{S_{\rm {Rb}}}$&projection of $S_{\rm {Rb}}$ on the space-fixed quantization axis\\
\end{tabular}
\end{ruledtabular}
\label{tab:quantumnumbers}
\end{table}

\vspace{0.5cm}

\newpage

\begin{table}
\caption{The positions ($B_0$) and widths ($\dB$) of $s$-wave resonances induced by an external electric field of 100~kV/cm for $^6$Li--$^{87}$Rb at magnetic fields below 2 kG. ($d$) denotes an $s$-wave electric-field-induced Feshbach resonance arising from a second order coupling through the $p$-wave channel to a $d$-wave closed channel state.  As a consequence, these resonances are exceedingly narrow.}
\begin{ruledtabular} 
\begin{tabular}{c c c c}
\multicolumn{2}{c}{Atomic States} & $B_0$ & $\dB$\\
$|f,m_f \rangle_{^6\rm Li}$ & $|f,m_f\rangle_{^{87}\rm Rb}$ & (G) & (G)\\
\hline
\hline
$|\frac{1}{2}, \frac{1}{2} \rangle$  & $|1, 1\rangle$ & $536.65$ $(d)$ & $ 0.01$\\
 & &$877.5$&$ 2.3$\\
 & &$654.52$&$ < 0.01$\\
\hline
$|\frac{1}{2}, \frac{1}{2} \rangle$ & $|1, 0\rangle$&$555.88$ $(d)$&$< 0.01$\\
 & &$885.8$&$2.6$\\
\hline

$|\frac{1}{2}, -\frac{1}{2} \rangle$ & $|1, 1\rangle$&$578.58$ $(d)$&$0.01$\\
 & &$707.70$&$<0.01$\\
 & &$770.50$&$<0.01$\\
\hline

$|\frac{1}{2}, -\frac{1}{2} \rangle$ & $|1, 0\rangle$&$596.01$ $(d)$&$< 0.01$\\
 & &$926.8$&$2.6$\\
\hline
$|\frac{3}{2}, \frac{3}{2} \rangle$ & $|1, -1\rangle$&$1242.5$&$12.7$\\
\end{tabular}
\end{ruledtabular}
\label{tab:resonances}
\end{table}

\newpage

\begin{figure}[ht]
\vspace{0.5cm}
\begin{center}
\includegraphics[scale=0.5]{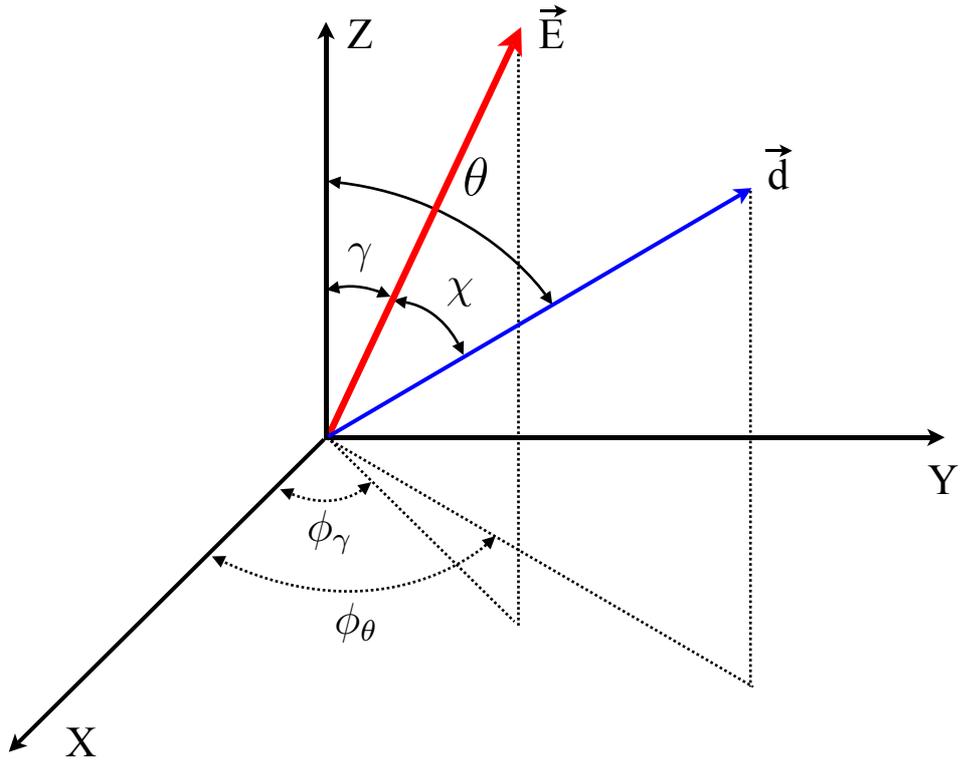}
\end{center}
\caption{(Color online) The coordinate system used for our calculations. $\vec{E}$ and $\vec{d}$ represent the vector of the external electric field and the dipole moment vector, respectively; $\gamma$ specifies the angle between the electric field vector and the quantization axis; $\theta$ is the angle between the dipole moment vector and the $z$-axis; and $\chi$ is the angle between $\vec{E}$ and $\vec{d}$.  The azimuthal angles, $\phi_\gamma$ and $\phi_\theta$, are measured from the positive $x$-axis to the orthogonal projection of the $\vec{E}$ and $\vec{d}$ vectors in the $x$-$y$ plane.}
\label{coordinate}
\end{figure}

\newpage

\begin{figure}[ht]
\vspace{0.5cm}
\begin{center}
\includegraphics[scale=0.5]{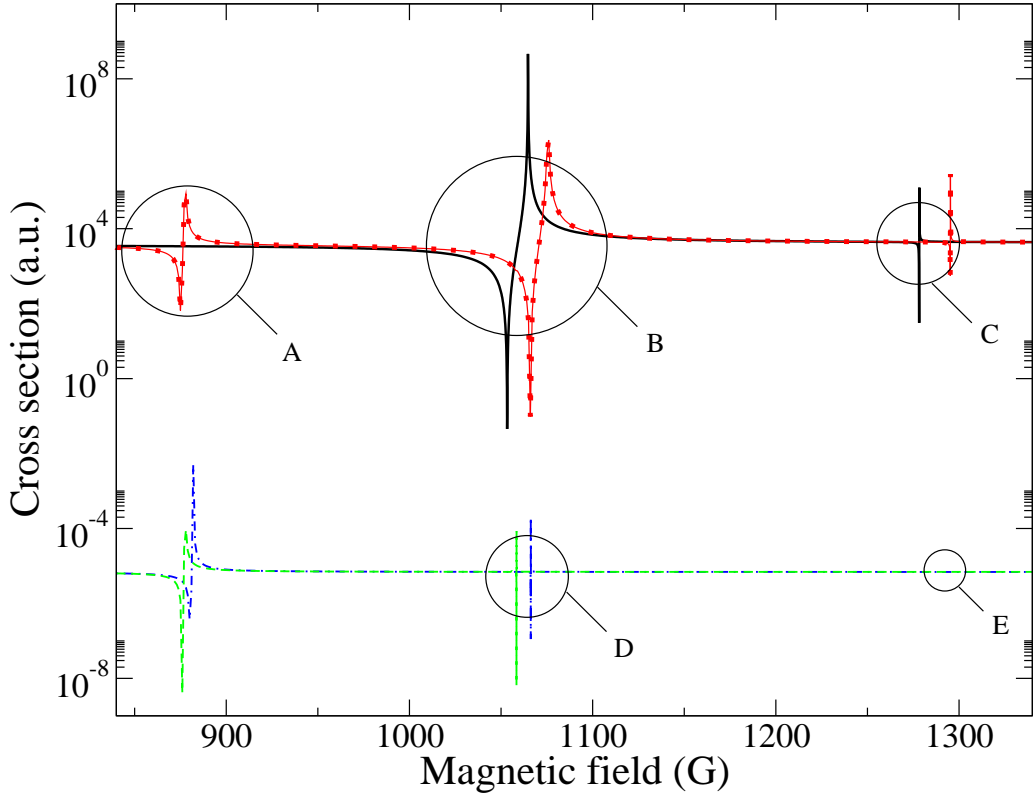}
\end{center}
\caption{(Color online) Magnetic field dependence of the elastic cross section for collisions between Li and Rb in the atomic spin state $|\frac{1}{2}, \frac{1}{2} \rangle_{^6\rm Li} \otimes |1, 1 \rangle_{^{87}\rm Rb}$.  These results were obtained for a collision energy of $10^{-7} \rm cm^{-1}$ and two different electric fields.  
The solid and dash-dotted curves show the $s$- and $p$-wave cross sections with $E=0$, while the dotted and dashed curves show the $s$- and $p$-wave cross sections when $E=100$~kV/cm.  Here, only the cross section for the $m_l=0$ state is shown for $p$-wave scattering.  At $A$ an $s$-wave resonance is induced by an intrinsic $p$-wave resonance.  Figure \ref{fig:s-wave-induced-resonance} shows this feature in more detail.  At $B$ and at $C$ an intrinsic $s$-wave resonance is shifted to higher magnetic fields (corresponding to a shift of the associated bound state to lower energy) due to the electric field coupling between bound states.  The observation that the shift of higher field resonances (e.g.~$C$) is typically larger than that of lower field resonances (e.g.~$B$) is discussed in the text.  At $D$ an intrinsic $p$-wave resonance is shifted to lower magnetic fields (corresponding to a shift of the associated bound state to higher energy).  At $E$ an induced $p$-wave resonance appears (invisible on this scale) due to the intrinsic $s$-wave resonance at $C$.}
\label{fig:ElectricFieldInducedWholePicture}
\end{figure}

\newpage

\begin{figure}[ht]
\vspace{0.5cm}
\begin{center}
\includegraphics[scale=0.5]{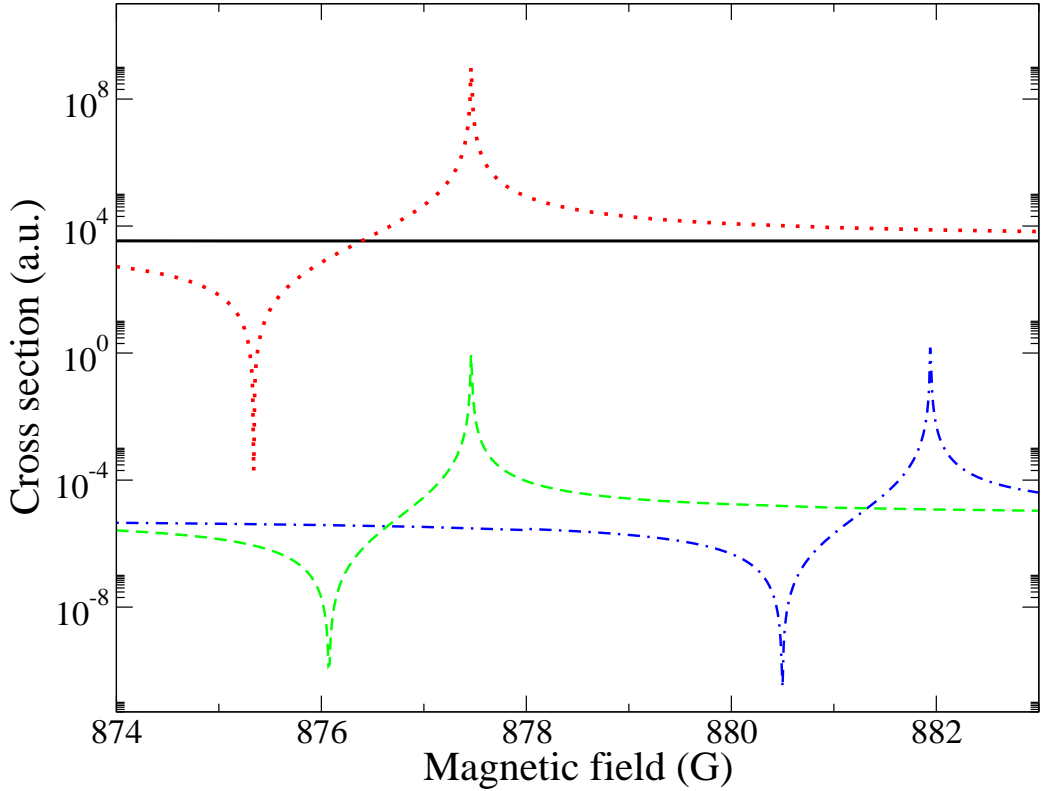}
\end{center}
\caption{(Color online) Magnetic field dependence of $s$- and $p$-wave elastic cross sections for atoms in the atomic spin state $|\frac{1}{2}, \frac{1}{2} \rangle_{^6\rm Li} \otimes |1, 1 \rangle_{^{87}\rm Rb}$ computed at different electric fields. This is the same feature at $A$ in Fig.~\ref{fig:ElectricFieldInducedWholePicture}.  The solid and dotted curves show the $s$-wave cross sections at $E=0$ and $E=100$~kV/cm, respectively.  The dot-dashed and dashed curves show the $p$-wave cross sections at $E=0$ and $E=100$~kV/cm, respectively.  This intrinsic $p$-wave resonance shifts to lower magnetic field (corresponding to the shift of the associated bound state to higher energy) as the electric field magnitude is increased.  The $s$-wave induced resonance appears at the same location as the intrinsic $p$-wave resonance, and its width grows with the strength of the electric field (see Fig.~\ref{fig:width-vs-E}).
Here only the cross section of the $m_l=0$ component is shown for the $p$-wave state is shown (Fig.~\ref{fig:p-splitting} shows the cross sections for all three components).  The collision energy is $10^{-7} \rm cm^{-1}$.}
\label{fig:s-wave-induced-resonance}
\end{figure}

\newpage

\begin{figure}[ht]
\vspace{0.5cm}
\begin{center}
\includegraphics[scale=0.5]{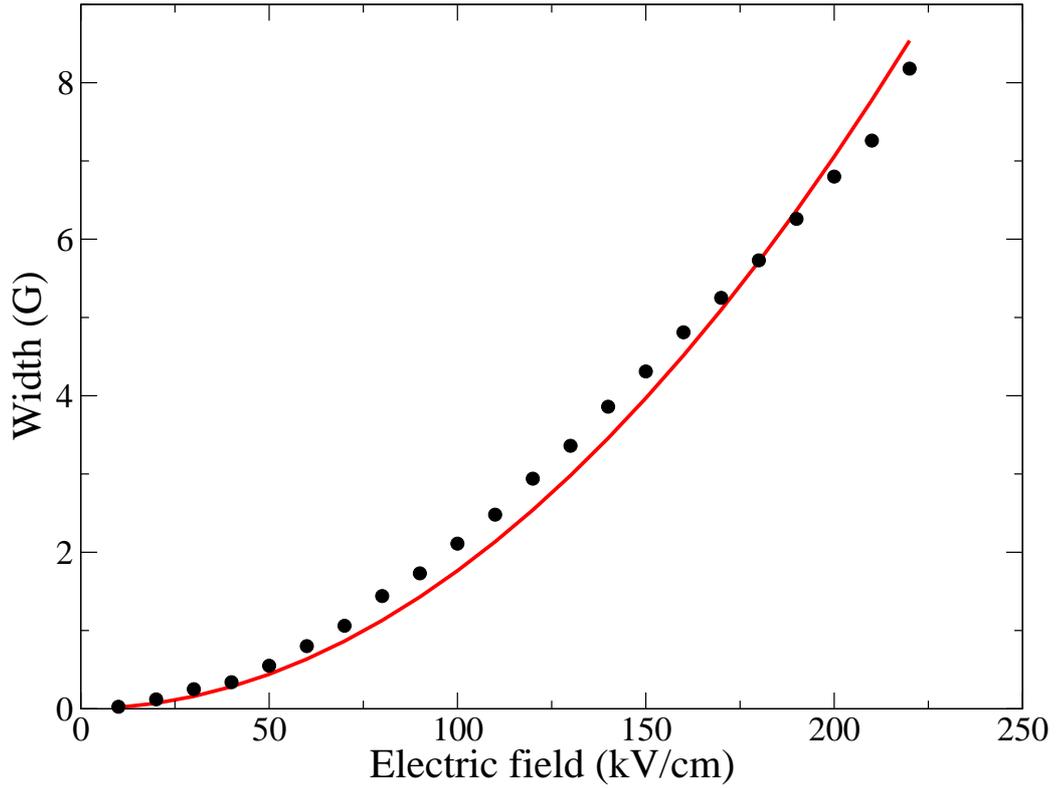}
\end{center}
\caption{(Color online) The width ($\Delta B$) of the $s$-wave electric-field-induced Feshbach resonance arising from the intrinsic $p$-wave resonance at 882~G as a function of the electric field magnitude.  Here $\gamma=0$ and the collision energy is $10^{-7} \rm cm^{-1}$.  The width appears to scale quadratically with $E$, at least for the electric fields below 200~kV/cm, and suggests that this induced resonance arises from an indirect coupling \cite{hemming:022705}.  The solid line is the fit $\Delta B = 1.76 \times 10^{-4} \; E^2$~G, where $E$ is in units of kV/cm.
}
\label{fig:width-vs-E}
\end{figure}

\newpage

\begin{figure}[ht]
\vspace{0.5cm}
\begin{center}
\includegraphics[scale=0.5]{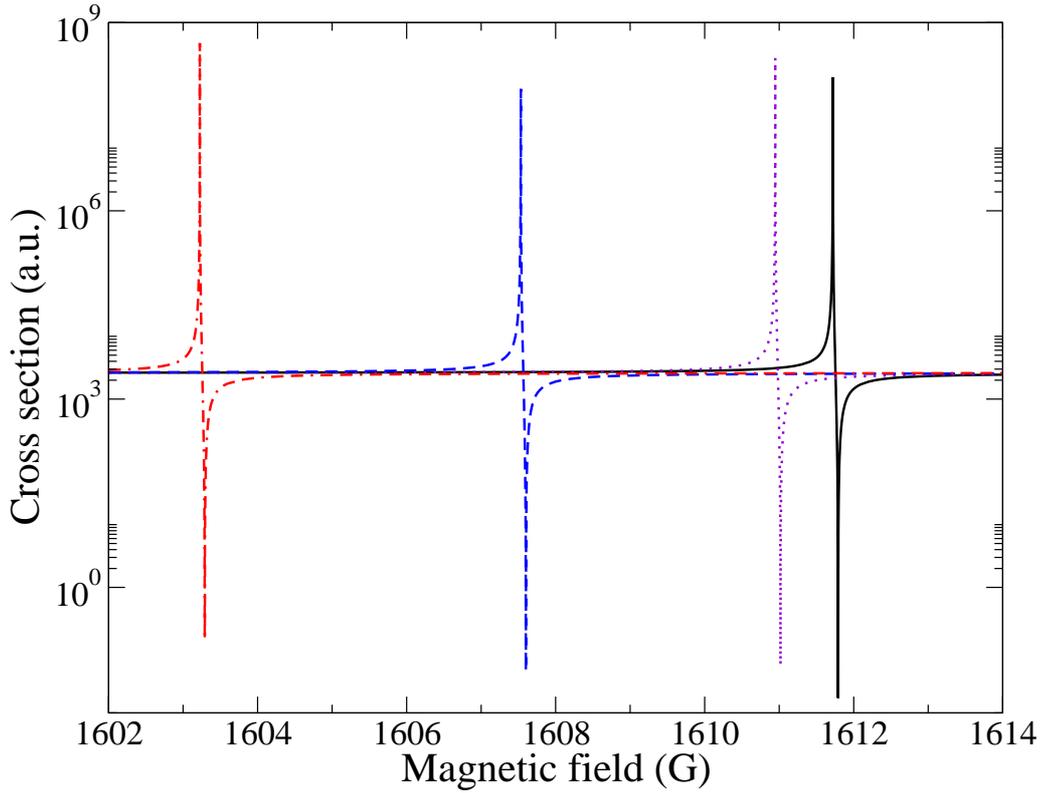}
\end{center}
\caption{(Color online) Magnetic field dependence of the $s$-wave elastic cross section for atoms in the atomic spin state $|\frac{1}{2}, -\frac{1}{2} \rangle_{^6\rm Li} \otimes |1, -1 \rangle_{^{87}\rm Rb}$ computed at different electric fields: $E = 0$~kV/cm (solid curve), $E = 30$~kV/cm (dotted curve), $E = 70$~kV/cm 
(dashed curve) and $E = 100$~kV/cm (dot-dashed curve).  An intrinsic $s$-wave resonance (whose position is 1611~G in the absence of an electric field) is observed to shift to lower magnetic fields as the electric field strength is increased.  Note: the shift direction is in the opposite sense to that of the intrinsic $s$-wave resonances in Fig.~\ref{fig:ElectricFieldInducedWholePicture}.  These results were obtained with a collision energy of $10^{-7} \rm cm^{-1}$.}
\label{fig:gradual-shift}
\end{figure}

\newpage

\begin{figure}[ht]
\vspace{0.5cm}
\begin{center}
\includegraphics[scale=0.5]{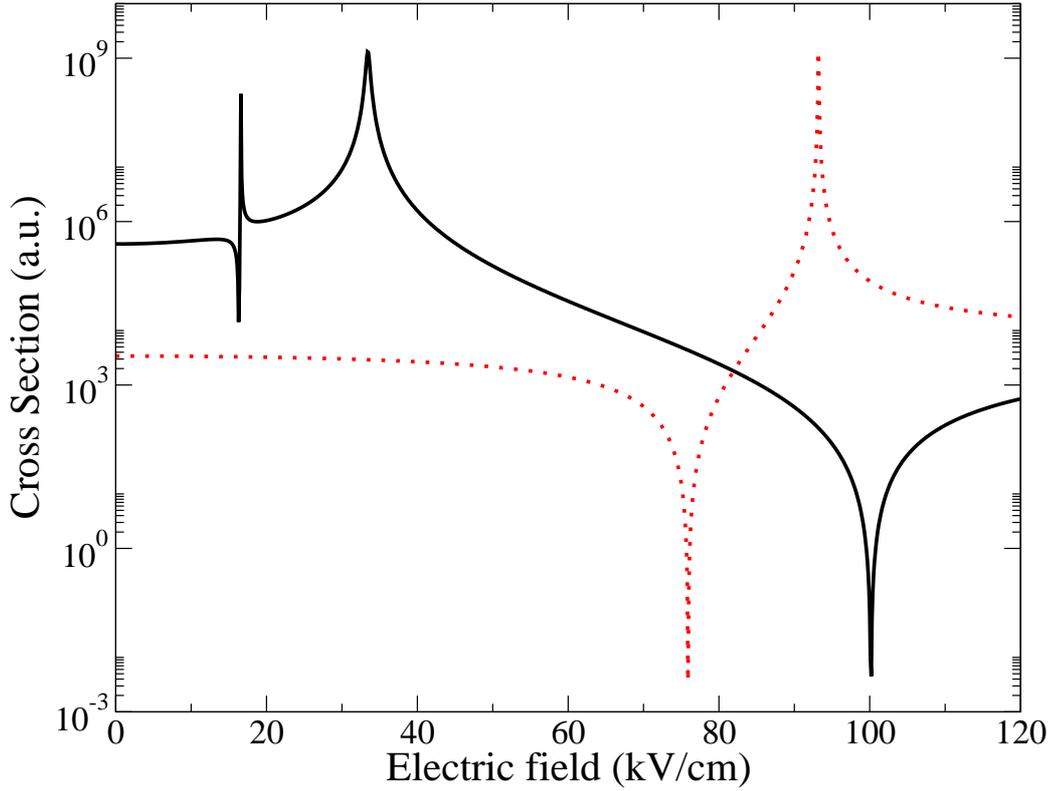}
\end{center}
\caption{(Color online) Variation of the cross section for $s$-wave collisions as a function of the electric field strength with the magnetic field fixed at 1066~G (solid line) and 878~G (dotted line) for atoms in the spin state $|\frac{1}{2}, \frac{1}{2} \rangle_{^6\rm Li} \otimes |1, 1 \rangle_{^{87}\rm Rb}$.  The large resonance feature shown in the solid curve is due to the shift of the intrinsic magnetic Feshbach resonance just below 1066~G to higher magnetic fields, while the small resonance feature at 16~kV/cm arises from the shift of an intrinsic $p$-wave resonance just above 1066~G to lower magnetic fields as the electric field increases.  The dotted curve shows a resonance feature associated with an electric-field-induced resonance (shown in Fig.~\ref{fig:s-wave-induced-resonance}) which moves from 882~G at $E=0$ down to a magnetic field below 877~G at $E=120$~kV/cm.  The collision energy is $10^{-7} \rm cm^{-1}$.}
\label{fig:CrossSectionVsElectricField}
\end{figure}

\newpage

\begin{figure}[ht]
\vspace{0.5cm}
\begin{center}
\includegraphics[scale=0.5]{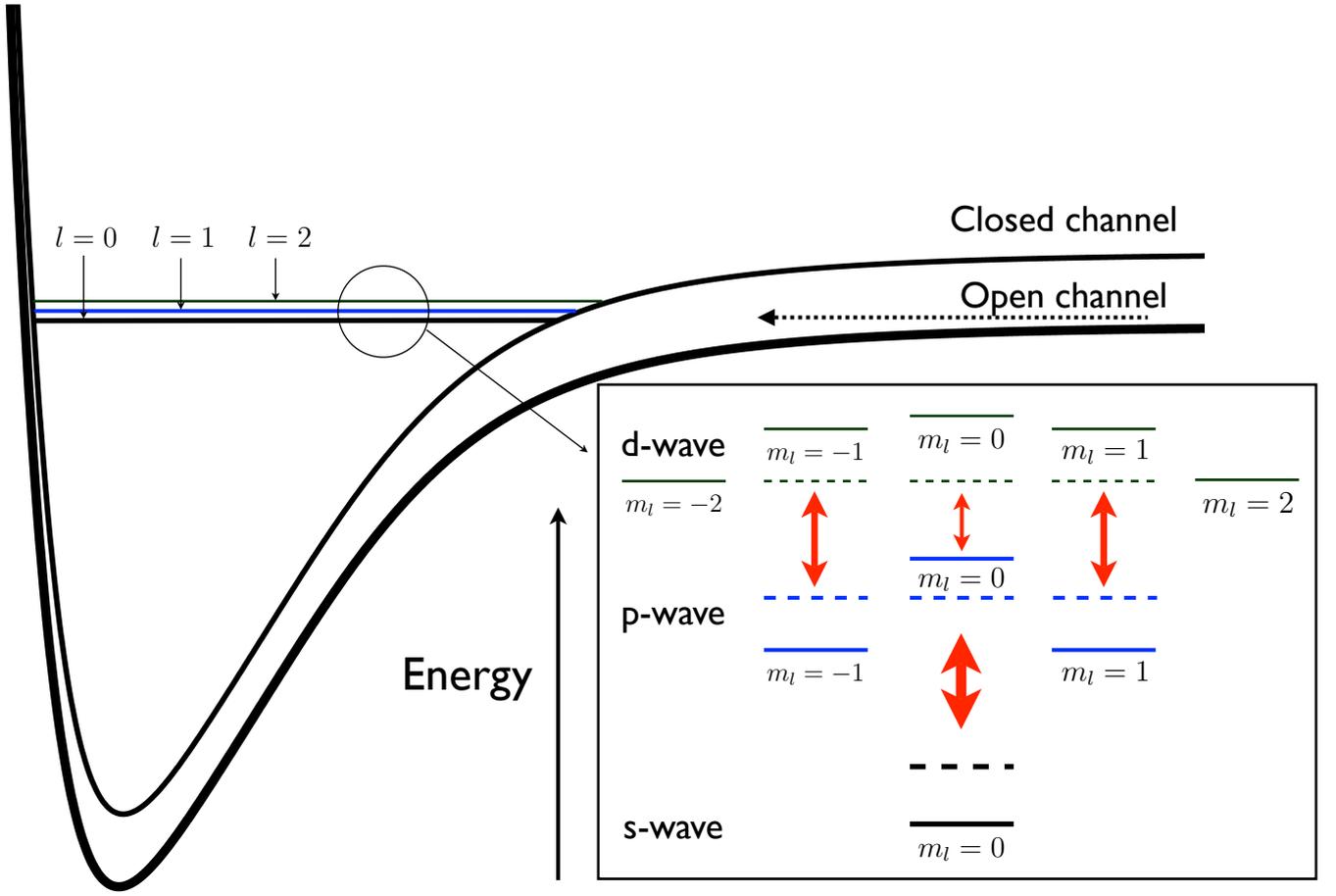}
\end{center}
\caption{(Color online) A schematic illustrating the mechanism of the splitting of $p$- and $d$-wave bound states resulting in the splitting of the corresponding Feshbach resonances.  For simplicity, only three adjacent bound state levels are shown.  The different partial wave potentials of each state are on this scale almost indistinguishable and are drawn here as a single potential.  The inset shows the energy levels associated with these three states.  The dotted lines indicate their energies in the absence of an electric field.  The coupling induced by the electric field is represented as double-ended arrows and shown for the case when the electric field is aligned along the magnetic field, i.e.~when $\gamma = 0$ states with the same $m_l$ value are coupled.  The coupling results in level repulsion and the new position of the states is indicated by the solid lines.  The degeneracy of the $p$- and $d$-wave bound states is broken and the associated Feshbach resonance splits into a multiplet with $l+1$ distinct resonances as shown in Figs.~\ref{fig:p-splitting} and \ref{fig:d-splitting}.  This simple picture predicts that the $s$-wave resonance should shift to higher magnetic fields (given the energy of the threshold moves down with increasing magnetic fields) and that the $m_l=0$ partial wave component should produce a new resonance at a magnetic field below the $m_l=1$ component - consistent with the motion of the resonances in 
Fig.~\ref{fig:ElectricFieldInducedWholePicture}
and
Fig.~\ref{fig:p-splitting}.  Of course, each state is coupled to all other bound states within the same spin manifold and with an orbital angular momenta differing by $\Delta l = \pm1$, resulting in splittings and shifts (e.g.~Fig.~\ref{fig:gradual-shift}) which may not follow the predictions of this simple picture
}
\label{fig:ResonanceRepulsionFig}
\end{figure}

\newpage

\begin{figure}[ht]
\vspace{0.5cm}
\begin{center}
\includegraphics[scale=0.5]{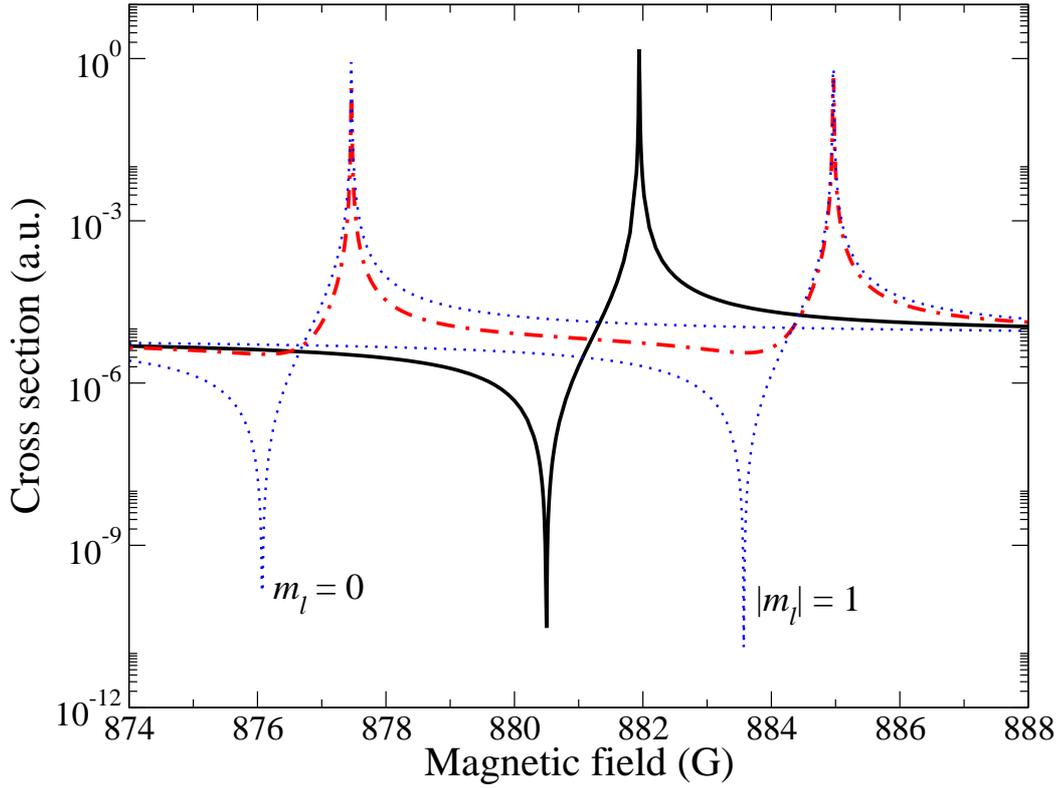}
\end{center}
\caption{(Color online) Magnetic field dependence of $p$-wave elastic cross section (averaged over all three orbital angular momentum components) for atoms in the atomic spin state $|\frac{1}{2}, \frac{1}{2} \rangle_{^6\rm Li} \otimes |1, 1 \rangle_{^{87}\rm Rb}$ computed at zero electric field (solid curve) and at $E = 100 \rm kV/cm$ (dot-dashed curve). The thin dotted curves show the magnetic field dependence of the cross section for the $|m_l| = 1$ and the $m = 0$ components separately.  The $p$-wave resonance splits into two distinct resonances, one occurring for the $m_l=0$ component and one for the $|m_l|=1$ components.  When the electric and magnetic fields are not co-linear, this segregation of the resonance multiplet breaks down as seen in Fig.~\ref{fig:different-gamma}.  The collision energy is $10^{-7} \rm cm^{-1}$.
}
\label{fig:p-splitting}
\end{figure}

\newpage

\begin{figure}[ht]
\vspace{0.5cm}
\begin{center}
\includegraphics[scale=0.5]{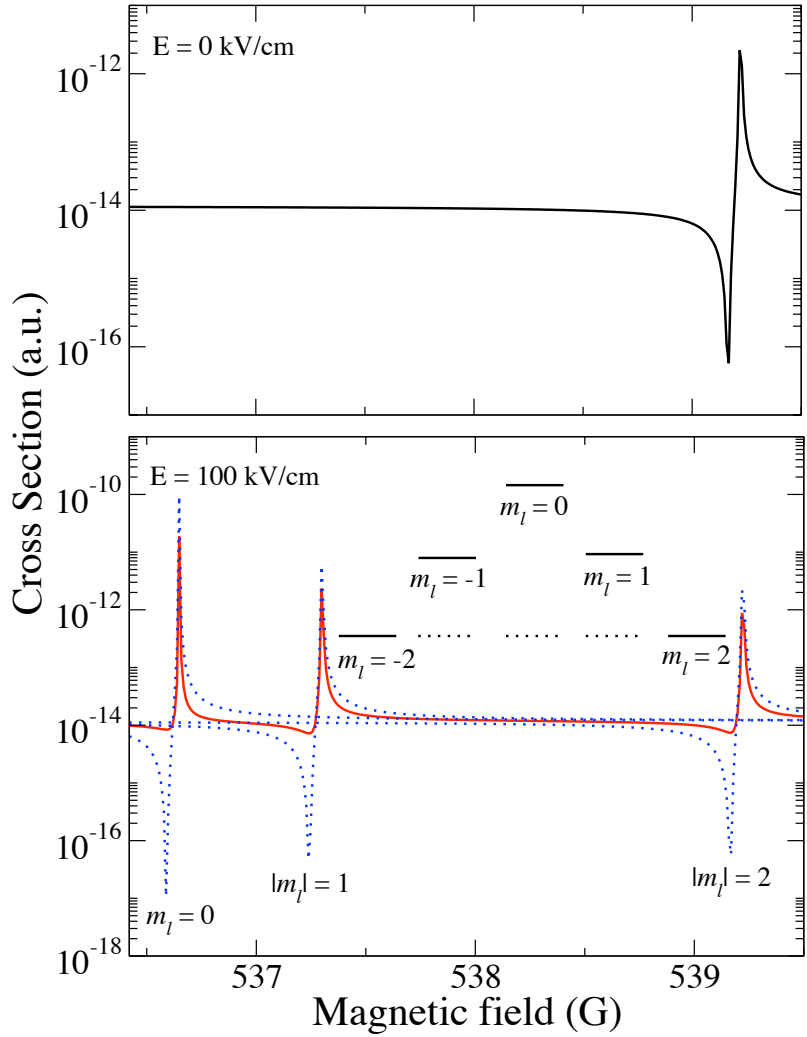}
\end{center}
\caption{(Color online) The upper panel shows the magnetic field dependence of the $d$-wave elastic cross section for atoms in the atomic spin state $|\frac{1}{2}, \frac{1}{2} \rangle_{^6\rm Li} \otimes |1, 1 \rangle_{^{87}\rm Rb}$ computed at zero electric fields (dotted-dashed curve).  The lower panel shows the magnetic field dependence of $d$-wave elastic cross section (solid curve). The contributions to the cross section from the $|m_l| = 2$,  $|m_l| = 1$ and the $m = 0$ components are shown (dotted curves) at $E = 100$~kV/cm. The $d$-wave resonance splits into $l+1=3$ distinct resonances corresponding to the splitting of the $d$-wave bound state levels drawn schematically in the lower panel.  The collision energy is $10^{-7} \rm cm^{-1}$.
}
\label{fig:d-splitting}
\end{figure}

\newpage

\begin{figure}[ht]
\vspace{0.5cm}
\begin{center}
\includegraphics[scale=0.5]{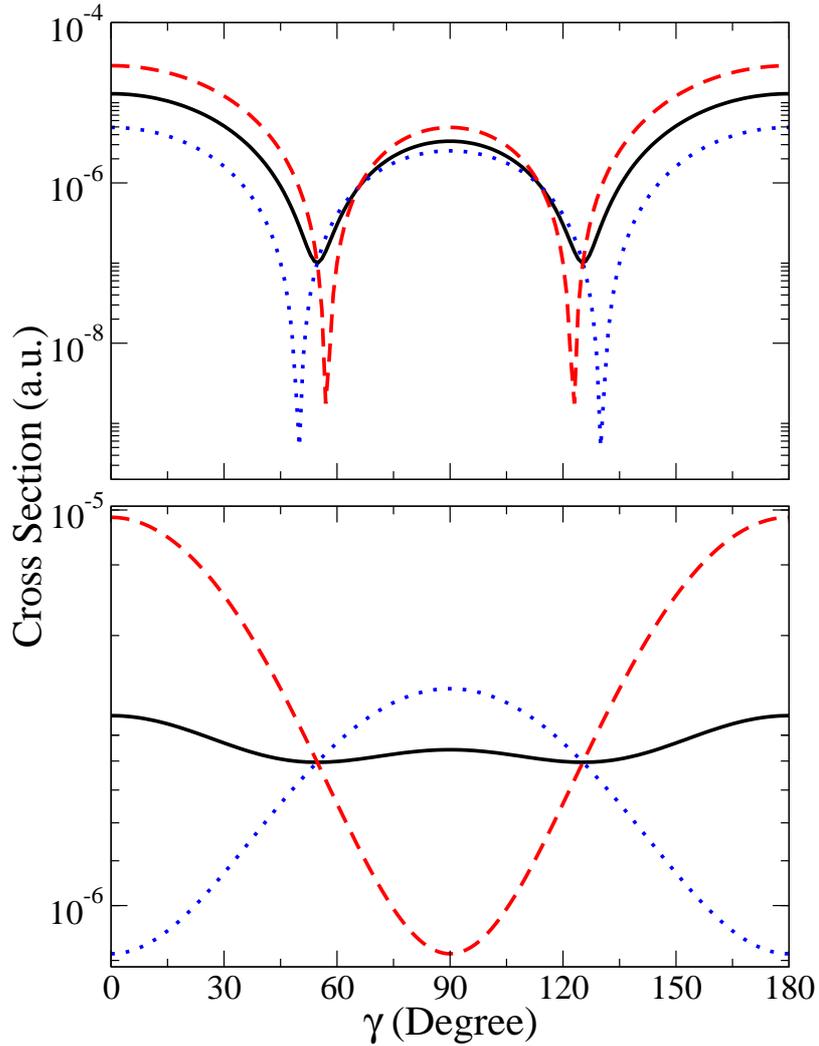}
\end{center}
\caption{(Color online) Total elastic cross section for different components of $p$-wave scattering versus the angle, $\gamma$, between the applied electric and magnetic fields.  The cross sections are shown for collisions in the $m_l=0$ state (dashed curve), the $|m_l|=1$ states (dotted curve), and the average (solid curve) of the cross sections over all three components for the atomic state $|\frac{1}{2}, \frac{1}{2} \rangle_{^6\rm Li} \otimes |1, 1 \rangle_{^{87}\rm Rb}$ and for $E = 100$~kV/cm .  The upper panel shows these cross sections at an applied magnetic field of 877.0~G which is near the resonance for the $m_l=0$ component while the lower panel is at a field of 881.9~G which is in between the resonances for the $m_l=0$ and $|m_l|=1$ components (see Fig.~\ref{fig:p-splitting}).  We observe that the shape of this variation changes dramatically near a resonance.  The collision energy is $10^{-7} \rm cm^{-1}$.
}
\label{fig:Erotation}
\end{figure}

\newpage

\begin{figure}[ht]
\vspace{0.5cm}
\begin{center}
\includegraphics[scale=0.5]{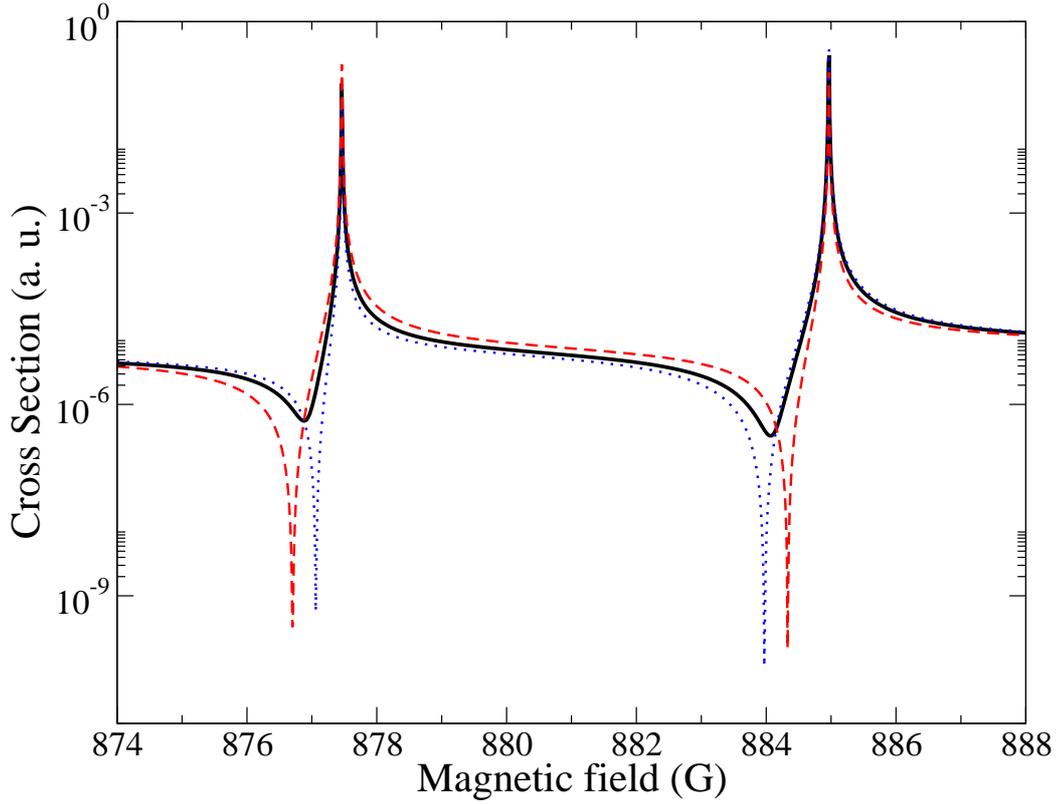}
\end{center}
\caption{(Color online) Magnetic field dependence of the elastic cross section for different components of $p$-wave scattering with an electric field, $E= 100$~kV/cm, titled with respect to the magnetic field axis by $\gamma=45^{\circ}$.  The cross sections are shown for collisions in the $m_l=0$ state (dashed curve), the $|m_l|=1$ states (dotted curve), and the average (solid curve) of the cross sections over all three components for the atomic state $|\frac{1}{2}, \frac{1}{2} \rangle_{^6\rm Li} \otimes |1, 1 \rangle_{^{87}\rm Rb}$.  The doublet structure of the $p$-wave resonance seen also in Fig.~\ref{fig:p-splitting} now appears on each of the three angular momentum projection components. The collision energy is $10^{-7} \rm cm^{-1}$.}
\label{fig:gamma45}
\end{figure}

\newpage

\begin{figure}[ht]
\vspace{0.5cm}
\begin{center}
\includegraphics[scale=0.5]{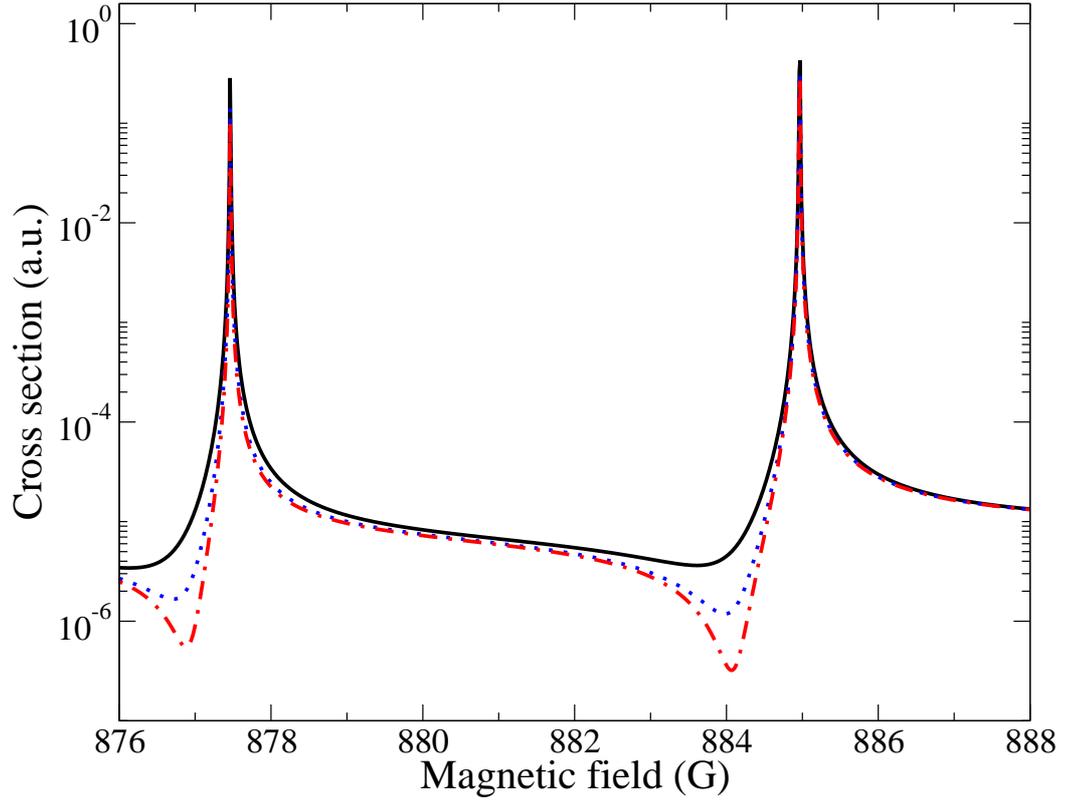} 
\end{center}
\caption{(Color online) Magnetic field dependence of elastic cross sections for atoms in the atomic spin state $|\frac{1}{2}, \frac{1}{2} \rangle_{^6\rm Li} \otimes |1, 1 \rangle_{^{87}\rm Rb}$ computed at $E = 100$~kV/cm with the orientation of the electric field at $\gamma$ = $0^{\circ}$ (solid curve), $45^{\circ}$ (dotted curve), and $90^{\circ}$ (dot-dashed curve). The collision energy is $10^{-7} \rm cm^{-1}$.}
\label{fig:different-gamma}
\end{figure}

\newpage

\clearpage

\newpage


\begin{thebibliography}{25}
\expandafter\ifx\csname natexlab\endcsname\relax\def\natexlab#1{#1}\fi
\expandafter\ifx\csname bibnamefont\endcsname\relax
  \def\bibnamefont#1{#1}\fi
\expandafter\ifx\csname bibfnamefont\endcsname\relax
  \def\bibfnamefont#1{#1}\fi
\expandafter\ifx\csname citenamefont\endcsname\relax
  \def\citenamefont#1{#1}\fi
\expandafter\ifx\csname url\endcsname\relax
  \def\url#1{\texttt{#1}}\fi
\expandafter\ifx\csname urlprefix\endcsname\relax\def\urlprefix{URL }\fi
\providecommand{\bibinfo}[2]{#2}
\providecommand{\eprint}[2][]{\url{#2}}

\bibitem[{\citenamefont{Tiesinga et~al.}(1993)\citenamefont{Tiesinga, Verhaar,
  and Stoof}}]{Tiesinga}
\bibinfo{author}{\bibfnamefont{E.}~\bibnamefont{Tiesinga}},
  \bibinfo{author}{\bibfnamefont{B.~J.} \bibnamefont{Verhaar}},
  \bibnamefont{and} \bibinfo{author}{\bibfnamefont{H.~T.~C.}
  \bibnamefont{Stoof}}, \bibinfo{journal}{Physical Review A}
  \textbf{\bibinfo{volume}{47}}, \bibinfo{pages}{4114} (\bibinfo{year}{1993}).

\bibitem[{\citenamefont{Weiner et~al.}(1999)\citenamefont{Weiner, Bagnato,
  Zilio, and Julienne}}]{Weiner}
\bibinfo{author}{\bibfnamefont{J.}~\bibnamefont{Weiner}},
  \bibinfo{author}{\bibfnamefont{V.~S.} \bibnamefont{Bagnato}},
  \bibinfo{author}{\bibfnamefont{S.}~\bibnamefont{Zilio}}, \bibnamefont{and}
  \bibinfo{author}{\bibfnamefont{P.~S.} \bibnamefont{Julienne}},
  \bibinfo{journal}{Reviews of Modern Physics} \textbf{\bibinfo{volume}{71}},
  \bibinfo{pages}{1} (\bibinfo{year}{1999}).

\bibitem[{\citenamefont{Kohler et~al.}(2006)\citenamefont{Kohler, Goral, and
  Julienne}}]{Kohler}
\bibinfo{author}{\bibfnamefont{T.}~\bibnamefont{Kohler}},
  \bibinfo{author}{\bibfnamefont{K.}~\bibnamefont{Goral}}, \bibnamefont{and}
  \bibinfo{author}{\bibfnamefont{P.~S.} \bibnamefont{Julienne}},
  \bibinfo{journal}{Reviews of Modern Physics} \textbf{\bibinfo{volume}{78}},
  \bibinfo{pages}{1311} (\bibinfo{year}{2006}).

\bibitem[{\citenamefont{Courteille et~al.}(1998)\citenamefont{Courteille,
  Freeland, Heinzen, van Abeelen, and Verhaar}}]{Courteille}
\bibinfo{author}{\bibfnamefont{P.}~\bibnamefont{Courteille}},
  \bibinfo{author}{\bibfnamefont{R.~S.} \bibnamefont{Freeland}},
  \bibinfo{author}{\bibfnamefont{D.~J.} \bibnamefont{Heinzen}},
  \bibinfo{author}{\bibfnamefont{F.~A.} \bibnamefont{van Abeelen}},
  \bibnamefont{and} \bibinfo{author}{\bibfnamefont{B.~J.}
  \bibnamefont{Verhaar}}, \bibinfo{journal}{Physical Review Letters}
  \textbf{\bibinfo{volume}{81}}, \bibinfo{pages}{69} (\bibinfo{year}{1998}).

\bibitem[{\citenamefont{Inouye et~al.}(1998)\citenamefont{Inouye, Andrews,
  Stenger, Miesner, Stamper-Kurn, and Ketterle}}]{Inouye}
\bibinfo{author}{\bibfnamefont{S.}~\bibnamefont{Inouye}},
  \bibinfo{author}{\bibfnamefont{M.~R.} \bibnamefont{Andrews}},
  \bibinfo{author}{\bibfnamefont{J.}~\bibnamefont{Stenger}},
  \bibinfo{author}{\bibfnamefont{H.~J.} \bibnamefont{Miesner}},
  \bibinfo{author}{\bibfnamefont{D.~M.} \bibnamefont{Stamper-Kurn}},
  \bibnamefont{and} \bibinfo{author}{\bibfnamefont{W.}~\bibnamefont{Ketterle}},
  \bibinfo{journal}{Nature} \textbf{\bibinfo{volume}{392}},
  \bibinfo{pages}{151} (\bibinfo{year}{1998}).

\bibitem[{\citenamefont{Leo et~al.}(2000)\citenamefont{Leo, Williams, and
  Julienne}}]{Leo}
\bibinfo{author}{\bibfnamefont{P.~J.} \bibnamefont{Leo}},
  \bibinfo{author}{\bibfnamefont{C.~J.} \bibnamefont{Williams}},
  \bibnamefont{and} \bibinfo{author}{\bibfnamefont{P.~S.}
  \bibnamefont{Julienne}}, \bibinfo{journal}{Physical Review Letters}
  \textbf{\bibinfo{volume}{85}}, \bibinfo{pages}{2721} (\bibinfo{year}{2000}).

\bibitem[{\citenamefont{Li et~al.}(2008)\citenamefont{Li, Singh, Tscherbul, and
  Madison}}]{Li}
\bibinfo{author}{\bibfnamefont{Z.}~\bibnamefont{Li}},
  \bibinfo{author}{\bibfnamefont{S.}~\bibnamefont{Singh}},
  \bibinfo{author}{\bibfnamefont{T.~V.} \bibnamefont{Tscherbul}},
  \bibnamefont{and} \bibinfo{author}{\bibfnamefont{K.~W.}
  \bibnamefont{Madison}}, \bibinfo{journal}{Physical Review A (Atomic,
  Molecular, and Optical Physics)} \textbf{\bibinfo{volume}{78}},
  \bibinfo{pages}{022710} (\bibinfo{year}{2008}).

\bibitem[{\citenamefont{Marzok et~al.}(2008)\citenamefont{Marzok, Deh,
  Zimmermann, Courteille, Tiemann, Vanne, and Saenz}}]{Marzok}
\bibinfo{author}{\bibfnamefont{C.}~\bibnamefont{Marzok}},
  \bibinfo{author}{\bibfnamefont{B.}~\bibnamefont{Deh}},
  \bibinfo{author}{\bibfnamefont{C.}~\bibnamefont{Zimmermann}},
  \bibinfo{author}{\bibfnamefont{P.~W.} \bibnamefont{Courteille}},
  \bibinfo{author}{\bibfnamefont{E.}~\bibnamefont{Tiemann}},
  \bibinfo{author}{\bibfnamefont{Y.~V.} \bibnamefont{Vanne}}, \bibnamefont{and}
  \bibinfo{author}{\bibfnamefont{A.}~\bibnamefont{Saenz}},
  \bibinfo{journal}{arXiv.org:cond-mat/0808.3967}  (\bibinfo{year}{2008}).

\bibitem[{\citenamefont{Danzl et~al.}(2008)\citenamefont{Danzl, Haller,
  Gustavsson, Mark, Hart, Bouloufa, Dulieu, Ritsch, and Nagerl}}]{Danzl}
\bibinfo{author}{\bibfnamefont{J.~G.} \bibnamefont{Danzl}},
  \bibinfo{author}{\bibfnamefont{E.}~\bibnamefont{Haller}},
  \bibinfo{author}{\bibfnamefont{M.}~\bibnamefont{Gustavsson}},
  \bibinfo{author}{\bibfnamefont{M.~J.} \bibnamefont{Mark}},
  \bibinfo{author}{\bibfnamefont{R.}~\bibnamefont{Hart}},
  \bibinfo{author}{\bibfnamefont{N.}~\bibnamefont{Bouloufa}},
  \bibinfo{author}{\bibfnamefont{O.}~\bibnamefont{Dulieu}},
  \bibinfo{author}{\bibfnamefont{H.}~\bibnamefont{Ritsch}}, \bibnamefont{and}
  \bibinfo{author}{\bibfnamefont{H.-C.} \bibnamefont{Nagerl}},
  \bibinfo{journal}{Science} \textbf{\bibinfo{volume}{321}},
  \bibinfo{pages}{1062} (\bibinfo{year}{2008}).

\bibitem[{\citenamefont{Ni et~al.}(2008)\citenamefont{Ni, Ospelkaus,
  de~Miranda, Pe'er, Neyenhuis, Zirbel, Kotochigova, Julienne, Jin, and
  Ye}}]{Ni}
\bibinfo{author}{\bibfnamefont{K.~K.} \bibnamefont{Ni}},
  \bibinfo{author}{\bibfnamefont{S.}~\bibnamefont{Ospelkaus}},
  \bibinfo{author}{\bibfnamefont{M.~H.~G.} \bibnamefont{de~Miranda}},
  \bibinfo{author}{\bibfnamefont{A.}~\bibnamefont{Pe'er}},
  \bibinfo{author}{\bibfnamefont{B.}~\bibnamefont{Neyenhuis}},
  \bibinfo{author}{\bibfnamefont{J.~J.} \bibnamefont{Zirbel}},
  \bibinfo{author}{\bibfnamefont{S.}~\bibnamefont{Kotochigova}},
  \bibinfo{author}{\bibfnamefont{P.~S.} \bibnamefont{Julienne}},
  \bibinfo{author}{\bibfnamefont{D.~S.} \bibnamefont{Jin}}, \bibnamefont{and}
  \bibinfo{author}{\bibfnamefont{J.}~\bibnamefont{Ye}},
  \bibinfo{journal}{Science} \textbf{\bibinfo{volume}{322}},
  \bibinfo{pages}{231} (\bibinfo{year}{2008}).

\bibitem[{\citenamefont{Krems}(2006)}]{Krems}
\bibinfo{author}{\bibfnamefont{R.~V.} \bibnamefont{Krems}},
  \bibinfo{journal}{Physical Review Letters} \textbf{\bibinfo{volume}{96}},
  \bibinfo{pages}{123202} (\bibinfo{year}{2006}).

\bibitem[{\citenamefont{Li and Krems}(2007)}]{Li2}
\bibinfo{author}{\bibfnamefont{Z.}~\bibnamefont{Li}} \bibnamefont{and}
  \bibinfo{author}{\bibfnamefont{R.~V.} \bibnamefont{Krems}},
  \bibinfo{journal}{Physical Review A (Atomic, Molecular, and Optical Physics)}
  \textbf{\bibinfo{volume}{75}}, \bibinfo{pages}{032709}
  (\bibinfo{year}{2007}).

\bibitem[{\citenamefont{Marinescu and You}(1998)}]{Marinescu}
\bibinfo{author}{\bibfnamefont{M.}~\bibnamefont{Marinescu}} \bibnamefont{and}
  \bibinfo{author}{\bibfnamefont{L.}~\bibnamefont{You}},
  \bibinfo{journal}{Physical Review Letters} \textbf{\bibinfo{volume}{81}},
  \bibinfo{pages}{4596} (\bibinfo{year}{1998}).

\bibitem[{\citenamefont{Melezhik and Hu}(2003)}]{Melezhik}
\bibinfo{author}{\bibfnamefont{V.~S.} \bibnamefont{Melezhik}} \bibnamefont{and}
  \bibinfo{author}{\bibfnamefont{C.-Y.} \bibnamefont{Hu}},
  \bibinfo{journal}{Physical Review Letters} \textbf{\bibinfo{volume}{90}},
  \bibinfo{pages}{083202} (\bibinfo{year}{2003}).

\bibitem[{\citenamefont{Marcelis et~al.}(2008)\citenamefont{Marcelis, Verhaar,
  and Kokkelmans}}]{Marcelis}
\bibinfo{author}{\bibfnamefont{B.}~\bibnamefont{Marcelis}},
  \bibinfo{author}{\bibfnamefont{B.}~\bibnamefont{Verhaar}}, \bibnamefont{and}
  \bibinfo{author}{\bibfnamefont{S.}~\bibnamefont{Kokkelmans}},
  \bibinfo{journal}{Physical Review Letters} \textbf{\bibinfo{volume}{100}},
  \bibinfo{pages}{153201} (\bibinfo{year}{2008}).

\bibitem[{\citenamefont{Silber et~al.}(2005)\citenamefont{Silber, Gunther,
  Marzok, Deh, Courteille, and Zimmermann}}]{Silber}
\bibinfo{author}{\bibfnamefont{C.}~\bibnamefont{Silber}},
  \bibinfo{author}{\bibfnamefont{S.}~\bibnamefont{Gunther}},
  \bibinfo{author}{\bibfnamefont{C.}~\bibnamefont{Marzok}},
  \bibinfo{author}{\bibfnamefont{B.}~\bibnamefont{Deh}},
  \bibinfo{author}{\bibfnamefont{P.~W.} \bibnamefont{Courteille}},
  \bibnamefont{and}
  \bibinfo{author}{\bibfnamefont{C.}~\bibnamefont{Zimmermann}},
  \bibinfo{journal}{Physical Review Letters} \textbf{\bibinfo{volume}{95}},
  \bibinfo{pages}{170408} (\bibinfo{year}{2005}).

\bibitem[{\citenamefont{Deh et~al.}(2008)\citenamefont{Deh, Marzok, Zimmermann,
  and Courteille}}]{Deh}
\bibinfo{author}{\bibfnamefont{B.}~\bibnamefont{Deh}},
  \bibinfo{author}{\bibfnamefont{C.}~\bibnamefont{Marzok}},
  \bibinfo{author}{\bibfnamefont{C.}~\bibnamefont{Zimmermann}},
  \bibnamefont{and} \bibinfo{author}{\bibfnamefont{P.~W.}
  \bibnamefont{Courteille}}, \bibinfo{journal}{Physical Review A (Atomic,
  Molecular, and Optical Physics)} \textbf{\bibinfo{volume}{77}},
  \bibinfo{pages}{010701} (\bibinfo{year}{2008}).

\bibitem[{\citenamefont{Bijlsma et~al.}(2000)\citenamefont{Bijlsma, Heringa,
  and Stoof}}]{Bijlsma}
\bibinfo{author}{\bibfnamefont{M.~J.} \bibnamefont{Bijlsma}},
  \bibinfo{author}{\bibfnamefont{B.~A.} \bibnamefont{Heringa}},
  \bibnamefont{and} \bibinfo{author}{\bibfnamefont{H.~T.~C.}
  \bibnamefont{Stoof}}, \bibinfo{journal}{Physical Review A}
  \textbf{\bibinfo{volume}{61}}, \bibinfo{pages}{053601}
  (\bibinfo{year}{2000}).

\bibitem[{\citenamefont{Aymar and Dulieu}(2005)}]{Aymar}
\bibinfo{author}{\bibfnamefont{M.}~\bibnamefont{Aymar}} \bibnamefont{and}
  \bibinfo{author}{\bibfnamefont{O.}~\bibnamefont{Dulieu}},
  \bibinfo{journal}{The Journal of Chemical Physics}
  \textbf{\bibinfo{volume}{122}}, \bibinfo{pages}{204302}
  (\bibinfo{year}{2005}).

\bibitem[{\citenamefont{Regal et~al.}(2003)\citenamefont{Regal, Ticknor, Bohn,
  and Jin}}]{Regal}
\bibinfo{author}{\bibfnamefont{C.~A.} \bibnamefont{Regal}},
  \bibinfo{author}{\bibfnamefont{C.}~\bibnamefont{Ticknor}},
  \bibinfo{author}{\bibfnamefont{J.~L.} \bibnamefont{Bohn}}, \bibnamefont{and}
  \bibinfo{author}{\bibfnamefont{D.~S.} \bibnamefont{Jin}},
  \bibinfo{journal}{Physical Review Letters} \textbf{\bibinfo{volume}{90}},
  \bibinfo{pages}{053201} (\bibinfo{year}{2003}).

\bibitem[{\citenamefont{Ticknor et~al.}(2004)\citenamefont{Ticknor, Regal, Jin,
  and Bohn}}]{Ticknor}
\bibinfo{author}{\bibfnamefont{C.}~\bibnamefont{Ticknor}},
  \bibinfo{author}{\bibfnamefont{C.~A.} \bibnamefont{Regal}},
  \bibinfo{author}{\bibfnamefont{D.~S.} \bibnamefont{Jin}}, \bibnamefont{and}
  \bibinfo{author}{\bibfnamefont{J.~L.} \bibnamefont{Bohn}},
  \bibinfo{journal}{Physical Review A} \textbf{\bibinfo{volume}{69}},
  \bibinfo{pages}{042712} (\bibinfo{year}{2004}).

\bibitem[{\citenamefont{Krems and Dalgarno}(2004)}]{Krems3}
\bibinfo{author}{\bibfnamefont{R.~V.} \bibnamefont{Krems}} \bibnamefont{and}
  \bibinfo{author}{\bibfnamefont{A.}~\bibnamefont{Dalgarno}},
  \bibinfo{journal}{The Journal of Chemical Physics}
  \textbf{\bibinfo{volume}{120}}, \bibinfo{pages}{2296} (\bibinfo{year}{2004}).

\bibitem[{\citenamefont{Hemming and Krems}(2008)}]{hemming:022705}
\bibinfo{author}{\bibfnamefont{C.~J.} \bibnamefont{Hemming}} \bibnamefont{and}
  \bibinfo{author}{\bibfnamefont{R.~V.} \bibnamefont{Krems}},
  \bibinfo{journal}{Physical Review A (Atomic, Molecular, and Optical Physics)}
  \textbf{\bibinfo{volume}{77}}, \bibinfo{eid}{022705}
  (pages~\bibinfo{numpages}{7}) (\bibinfo{year}{2008}).

\bibitem[{\citenamefont{Tsuei and Kirtley}(2000)}]{Tsuei}
\bibinfo{author}{\bibfnamefont{C.~C.} \bibnamefont{Tsuei}} \bibnamefont{and}
  \bibinfo{author}{\bibfnamefont{J.~R.} \bibnamefont{Kirtley}},
  \bibinfo{journal}{Physical Review Letters} \textbf{\bibinfo{volume}{85}},
  \bibinfo{pages}{182} (\bibinfo{year}{2000}).

\bibitem[{\citenamefont{Zare}(1988)}]{Zare}
\bibinfo{author}{\bibfnamefont{R.}~\bibnamefont{Zare}},
  \emph{\bibinfo{title}{Angular Momentum}} (\bibinfo{publisher}{John Wiley \&
  Sons}, \bibinfo{year}{1988}).

\end{thebibliography}


\end{document}